\documentclass[12pt, preprint]{aastex}

\usepackage{amsmath}
\usepackage{graphics}
\def\ngc#1{\hbox{NGC\,#1}}

\def\B{$B$}
\def\V{$V$}

\def\bmv{\hbox{\it B--V\/}}

\def\Sec{${}^{\prime\prime}$\llap{.}}
\def\deg{${}^\circ$}
\def\min{${}^{\prime}$}
\def\sec{${}^{\prime\prime}$}
\def\hr{${}^{\rm\scriptstyle h}$}
\def\tm{${}^{\rm\scriptstyle m}$}

\def\Ts{${}^{\rm\scriptstyle s}$\llap{.}}
\def\gsim{\;\lower.6ex\hbox{$\sim$}\kern-9.5pt\raise.4ex\hbox{$>$}\;}
\def\lsim{\;\lower.6ex\hbox{$\sim$}\kern-9.2pt\raise.4ex\hbox{$<$}\;}

\def\cf{{\it cf.}}

\def\leoi{Leo~I\/}

\shorttitle{Variable Stars in Leo~I}
\shortauthors{Stetson et al.}

\begin{document}
\title{Homogeneous Photometry VI: Variable Stars in the Leo~I Dwarf Spheroidal Galaxy${}^*$}

\author{
Peter~B.~Stetson\altaffilmark{1}, 
Giuliana~Fiorentino\altaffilmark{2},
Giuseppe~Bono\altaffilmark{3,4},
Edouard~J.~Bernard\altaffilmark{5},
Matteo~Monelli\altaffilmark{6,7},
Giacinto~Iannicola\altaffilmark{4},
Carme~Gallart\altaffilmark{6,7},
and Ivan~Ferraro\altaffilmark{4}
}

\altaffiltext{1}{Dominion Astrophysical Observatory, NRC-Herzberg, National Research Council, 5071 West Saanich Road, Victoria, BC V9E 2E7, Canada}
\altaffiltext{2}{INAF-Osservatorio Astronomico di Bologna, via Ranzani 1, 40127, Bologna, Italy}
\altaffiltext{3}{Dipartimento di Fisica, Universit\`{a} di Roma Tor Vergata, via della Ricerca Scientifica 1, 00133 Rome, Italy}
\altaffiltext{4}{INAF-Osservatorio Astronomico di Roma, via Frascati 33, Monte Porzio Catone, Rome, Italy}
\altaffiltext{5}{SUPA, Institute for Astronomy, University of Edinburgh, Royal Observatory, Blackford Hill, Edinburgh EH9 3HJ, UK}
\altaffiltext{6}{Instituto de Astrof\'{i}sica de Canarias, Calle Via Lactea, E38200 La Laguna, Tenerife, Spain}
\altaffiltext{7}{Departamento de Astrof\'{i}sica, Universidad de La Laguna, Tenerife, Spain}
\date{\centering drafted \today\ / Received / Accepted }

\begin{abstract}

From archival ground-based images of the Leo~I dwarf spheroidal
galaxy we have identified and characterized the pulsation
properties of 164 candidate RR Lyrae variables and 55 candidate
Anomalous and/or short-period Cepheids.  We have
also identified nineteen candidate long-period variable stars and
thirteen other candidate variables whose physical nature is
unclear, but due to the limitations of our observational material
we are unable to estimate reliable periods for them. On the basis
of its RR Lyrae star population Leo~I is confirmed to be an
Oosterhoff-intermediate type galaxy, like several other dwarf
spheroidals. From the RR Lyrae stars we have derived a range of
possible distance moduli for \leoi: $22.06\pm0.08 \lsim \mu_0
\lsim 22.25\pm0.07\,$mag depending on the metallicity assumed for
the old population ([Fe/H] from --1.43 to --2.15). This is in
agreement with previous independent estimates.  We show that in
their pulsation properties, the RR Lyrae stars---representing the
oldest stellar population in the galaxy---are not significantly
different from those of five other nearby, isolated dwarf
spheroidal galaxies.  A similar result is obtained when comparing
them to RR Lyrae stars in recently discovered ultra-faint dwarf
galaxies.  We are able to compare the period distributions and
period-amplitude relations for a statistically significant sample
of ab--type RR Lyrae stars in dwarf galaxies ($\sim\,$1300 stars)
with those in the Galactic halo field ($\sim\,$14,000 stars) and
globular clusters ($\sim\,$1000 stars).  Field RRLs show a
significant change in their period distribution when moving from
the inner ($d_G \lsim\,$14 kpc) to the outer ($d_G \gsim\,$14 kpc)
halo regions. This suggests that the halo formed from (at least)
two dissimilar progenitors or types of progenitor. Considered
together, the RR Lyrae stars in classical dwarf spheroidal and
ultra-faint dwarf galaxies---as observed today---do not appear to
follow the well defined pulsation properties shown by those in
either the inner or the outer Galactic halo, nor do they have the
same properties as RR Lyraes in globular clusters. In particular,
the samples of fundamental-mode RR Lyrae stars in dwarf galaxies
seem to lack {\it H\/}igh {\it A\/}mplitudes and {\it S\/}hort
{\it P\/}eriods (``HASP'': $A_V \ge 1.0\,$mag and
P$\,\lsim\,$0.48$\,$d) when compared with those observed in the
Galactic halo field and globular clusters.  The observed
properties of RR Lyrae stars do not support the idea that
currently existing classical dwarf spheroidal and ultra-faint
dwarf galaxies are surviving {\it representative\/} examples of
the original building blocks of the Galactic halo.  

\end{abstract}


\maketitle

%%%%%%%%%%%%%%%%%%%%%%%%%%%%%%%%%%%%%%%%%%%%%%%%%%%%%%%%%%%%%%%%%%%%

${}^*${\footnotesize Based in part on data obtained through the
facilities of the Canadian Astronomy Data Centre operated by the
National Research Council of Canada with the support of the
Canadian Space Agency; data obtained from the ESO Science Archive
Facility under muliple requests by the authors; data obtained from
the Isaac Newton Group Archive, which is maintained as part of the
CASU Astronomical Data Centre at the Institute of Astronomy,
Cambridge; and data distributed by the NOAO Science Archive. NOAO
is operated by the Association of Universities for Research in
Astronomy (AURA) under cooperative agreement with the National
Science Foundation.}

\section{Introduction}

The nearby dwarf spheroidal galaxies (dSphs,
L$\gsim$10$^5$L$_{\odot}$) are important for understanding the
formation and evolution of the Milky Way Galaxy.  Their very old ages ($\gsim$10 Gyr)
and low mean metallicities ([Fe/H]$\lsim$--1.3 dex) have long been
used as evidence for the idea that these systems are fossils
from the early Universe:  surviving examples of the pre-galactic
units that formed the Galactic stellar halo, as
predicted by hierarchical models of galaxy formation
\citep[see][and references therein]{white91,salvadori09}. This view has been
widely held for at least the last twenty years, since
detailed stellar population studies have been made possible
by new deep color-magnitude diagrams (CMDs) and medium- and
high-resolution spectroscopic analyses \citep{mateo98,tolstoy09}.
Today, we have at our disposal accurate star-formation histories
\citep{tolstoy96,gallart99,cole07,monelli10b,monelli10a,hidalgo11,deboer12a,deboer12b} 
and chemical abundance distributions 
\citep{venn04,kirby09,kirby11,fabrizio12,lemasle12,kirby13} 
for a large number of nearby dSphs and dwarf irregulars (dIs).
However, high-resolution spectroscopy has shown that the [$\alpha$/Fe]
abundance patterns of stars in dSphs do not seem to resemble those
of Galactic halo stars \citep{shetrone01,venn04,pritzl05b}.  This
could pose some problems for their identification as
representative building blocks of the halo. On the other hand, it could
also be explained by the fact that present-day dSphs, having {\it
not\/} accreted at early epochs, have had a Hubble time to evolve as
independent entities under conditions that are different in a variety of ways from those experienced by
field halo stars.  In support of this interpretation,
\citet{font06} have argued that differences in the chemical
abundances of the Galactic halo and its satellites can be
explained within their hierarchical formation model of the halo,
which carefully tracks the orbital evolution and tidal disruption
of its satellites.

A new source of puzzlement was raised by \citet{helmi06}, who
obtained large samples of low- and intermediate-resolution spectra
of red giant branch (RGB) stars of the Sculptor, Sextans, Fornax,
and Carina dSph galaxies as part of a huge observational effort
devoted to the characterization of the chemical abundances of
those systems \citep[see][and references therein]{tolstoy06}. For the first time,
\citet{helmi06} noticed a significant lack of extremely metal-poor
stars ([Fe/H]$\lsim$--3 dex) in the galaxies' metallicity
distributions when compared with that of the Galactic halo. The
most metal-poor stars in a galaxy are likely to be also the oldest
ones so---taken at face value---this evidence again seems to argue
against the idea that nearby dSphs are surviving examples of the
main contributors to the Milky Way halo. In order to further
investigate this issue, \citet{starkenburg10} revisited the
calibration of [Fe/H] as inferred from the observed strength of
the infrared calcium triplet, which has been widely used to
estimate the metal content of large samples of RGB stars in nearby
dSph galaxies. The initial calibration of this relationship had
presumed that it was linear over the relevant range of abundance,
but in the new study it turned out to be nonlinear for
[Fe/H]$\lsim$--2 dex, thus modifying the shape of the metallicity
distribution in the metal-poor tail. Using the new calibration,
Starkenburg et al.\ suggested that classical dSphs are not as
devoid of extremely low-metallicity stars as previously believed.
Selecting those stars with [Fe/H]$\lsim$--3 dex based on this new
calibration of the IR calcium triplet, \citet{starkenburg13a}
followed up with high-resolution spectroscopy that confirmed the stars'
extremely low metallicities.  Today more than 30 extremely
metal-poor stars have been identified in dSphs, and their chemical
abundance patterns now seem to more closely resemble those in the
Galactic halo \citep[e.g.,][]{kirby08,koch08,kirby09,frebel10a,frebel10b,starkenburg13a}. 

Furthermore, the discovery of ultra-faint dwarf galaxies
\citep[UFDs,
L$\sim$10$^3$-10$^5$L$_{\odot}$][]{willman06,zucker06,belokurov06}
seems to shed some more light on this intriguing scenario. They
are numerous and very old ($\gsim\,$10 Gyr), and they contain extremely
metal-poor stars \citep{kirby08}.  Thus to some investigators they
show promise as ideal candidates for the original building blocks of
the Galactic halo, that may at the same time help solve the
missing-satellite problem \citep[see][and references
therein]{white91,salvadori09}. Moreover, spectroscopic investigations of several UFDs provide strong
constraints on the dynamical interaction between their baryonic
content and their dark matter halos \citep{mayer10,gilmore13}

Tracing the oldest stellar component in a galaxy, RR Lyrae
variable stars (RRLs) have the potential to independently test
this interesting global picture through their pulsation
properties. In particular, RRLs observed in both the Galactic halo
and globular clusters (GCs) show a well defined dichotomy in their
period distribution \citep[see discussion in][and references
therein]{contreras13}. This behavior was first observed by
\citet{oosterhoff39} in Galactic GCs:  he divided them into two
groups according to the mean period of their RRab variable stars. 
Specifically, Oosterhoff~I (OoI) GCs show mean periods around
$\left< \hbox{\rm Pab} \right>\sim 0.55\,$d, whereas OoII GCs show longer periods
($\left< \hbox{\rm Pab} \right>\sim 0.65\,$d).  Further observations disclosed that
the mean periods of the RRc-type variables are also
different---again shorter in OoI clusters---and also that the
fraction of RRc variables among RR~Lyraes of all types is
systematically smaller in OoI ($\sim\,$17\%) than in OoII
($\sim\,$44\%). Many studies have also demonstrated a relationship
between the metallicity of the host cluster and the Oosterhoff
type of its variable-star population: OoII clusters are typically
very metal-poor \hbox{([Fe/H]$\,\sim\,$--2 dex)} halo objects, while OoI
clusters cover a broader range of generally higher metallicities.  At
present---75 years later---there is still no full
and satisfying explanation for this phenomenology among the GCs.  

More recently, major observational surveys have shown that 
field RRLs in the Milky Way halo show the same period
dichotomy as those in the globulars (\citealt{bono97b}; ASAS, \citealt{pojmanski05}; LONEOS:
\citealt{miceli08}; LINEAR: \citealt{sesar13}, etc.). In contrast, those 
in nearby Local Group galaxies and their GCs are perversely characterized by
mean periods precisely in the range $0.58\,\hbox{\rm d} < \left< \hbox{\rm Pab} \right><0.62\,$d---the
so-called ``Oosterhoff gap''---that is avoided by the Milky Way
halo cluster and field populations \citep[see][for a detailed
discussion]{bono94,catelan09}. However this comparison, based on the
statistical properties of very different samples of stars that may
be subject to different selection biases, is not truly decisive
in characterizing the differences between GCs, the Galactic halo,
and dwarf galaxies.  

Recently, extensive variability searches to characterize the RRL
populations in UFDs have been carried out
\citep{siegel06,dallora06,kuehn08,greco+08,moretti09,musella09,
dallora12,musella12,clementini12,garofalo13,boettcher13}.  Given
the very small sample of RRL stars observed in most of the cases,
it is very difficult even to provide a clear Oosterhoff
classification for most UFDs. A few exceptions do exist: Bootes
\citep[11 RRLs, see][]{dallora06} is the sole example of an
apparently well-established OoII type, whereas Ursa Major I
\citep[7 RRLs, see][]{garofalo13} is classified as
Oo-intermediate, resembling the classical dSph galaxies.  Indeed,
at the present date we might regard the distinction between UFDs
and dSphs as largely a matter of semantics:  according to the
available evidence, the two terms may merely refer to the lower-
and higher-mass members of a single family of objects \citep[see,
e.g.,][especially Figs.~6, 7, and 12]{mcconnachie12}.

This paper is part of the effort of one of us (PBS) to establish
and maintain a database of homogeneous photometry for resolved
star clusters and galaxies. This effort seeks the greatest
completeness and longest time span possible by exploiting public-domain data
currently available from astronomical archives.  In this paper we
analyse in detail the properties of variable stars that we have
identified in the Leo~I dwarf spheroidal galaxy.  Part of these
data have already been presented in \citet{fiorentino12d}, where
we focused our attention on understanding the intermediate-age
central-helium-burning stellar population (Anomalous and/or
short-period Cepheids, hereinafter AC/spCs). Here we present the
full infrastructure underlying that discussion:  identifications
and equatorial positions for the complete set of variable star candidates
together with their periods and light curve parameters.  In
\S\ref{rr} we describe in detail the RRLs
of Leo~I.  In \S\ref{cep} we perform a slightly more detailed analysis of the
AC/spC populations through mass determinations.  Next, in
\S\ref{dSph} and \S\ref{GCfield} we discuss the variable stars in Leo~I within the
larger context of Milky Way formation and evolution discussed
above. In particular we will compare the period distributions and
period-amplitude relations of the RRLs in (a)~our data for Leo~I
and literature data for other classical dSph galaxies with
literature data for (b)~the Galactic halo field, (c)~the
Galactic halo GCs, and (d)~the new Galactic halo UFDs. A
final summary discussion closes the main body of the paper.  An
Appendix addresses the reidentification of previously published
variable candidates in Leo~I.

\section{Observations}

The observational material for this study consists of 1,884
individual CCD images obtained on 48 nights during 32 observing
runs.  These data are contained within a much larger data
collection ($\sim\,$400,000 images, $\sim\,$500 observing runs)
compiled and maintained by the first author. 
Summary details of all 32 observing runs used here are given in
Table~\ref{obs}. In fact, the observing runs designated ``23
wfi4'' and ``24 wfi'' were the same run, as were those designated
``27 suba'' and ``28 suba2.''  However, in each case different
subsets of the images had been provided to us through
different channels.  To avoid any possible, unknown differences in
the ways the images had been treated before we obtained them, we
chose to keep them separate. Considering all the \leoi\ images
together, the median seeing for our observations was 1\Sec0
arcseconds; the 25th and 75th percentiles were 0\Sec8 and 1\Sec3
arcseconds; the 10th and 90th percentiles were 0\Sec7 and 1\Sec6
arcseconds.  

The photometric reductions were all carried out using standard
DAOPHOT and \hbox{ALLFRAME} procedures \citep{stetson87,stetson94}
to perform profile-fitting photometry, which was then referred
to a system of synthetic-aperture photometry by the method of
growth-curve analysis \citep{stetson90}.  There were insufficient
$U$-band observations from photometric occasions to establish a
reliable $U$ magnitude system in Leo~I, and we have made no
attempt to calibrate those data photometrically.  The few $U$-band
images that we do have, however, were included in the ALLFRAME
reductions to exploit whatever information they can provide on the
completeness of the star list and the precision of the astrometric
positions.  

Calibration of these instrumental data to the photometric system
of \cite{landolt92} \citep[see also][]{landolt73,landolt83} was
carried out as described by \citet{stetson00,stetson05a}. If we
define a ``dataset" as the accumulated observations from one CCD
chip on one night with photometric observing conditions, or one
chip on one or more nights from the same observing run with
non-photometric conditions, the data for \leoi\ were contained
within 95 different datasets, each of which was individually
calibrated to the Landolt system.  Of these 95 datasets, 63 were
obtained and calibrated as ``photometric,'' meaning that
photometric zero points, color transformations, and extinction
corrections were derived from all standard fields observed during
the course of each night, and applied to the \leoi\ observations. 
The other 32 datasets were reduced as ``non-photometric"; in this
case, color transformations were derived from all the standard
fields observed, but the photometric zero point of each individual
CCD image was derived from local \leoi\ photometric standards
contained within the image itself; these local standards were
established by us from the images obtained during photometric
conditions.  Considering the 95 datasets employed here, the median
values for the root-mean-square residuals of the observed results
for the standard stars relative to their established standard
values were 0.029$\,$mag in $B$ (minimum 0.007, maximum 0.079),
0.026 in $V$ (0.011, 0.076), 0.024 in $R$ (0.007, 0.059), and
0.024 in $I$ (0.011, 0.090).  The (minimum, median, maximum)
numbers of standard-star observations contained in each dataset
were (88, 1644, 15084) in $B$, (79, 2059, 25664) in $V$, (85,
1440, 8742) in $R$, and (26, 1526, 14790) in $I$.

The different cameras employed projected to different areas on the
sky, and of course the telescope pointings differed among the
various exposures.  Furthermore, the Wide-Field Camera on the
Isaac Newton Telescope, the Wide-Field Imager on the 2.2m MPG/ESO
telescope on La Silla, Suprime-Cam on the Subaru Telescope, FORS
on the VLT on Paranal, and LBC on the Large Binocular Telescope on
Mount Graham consist of mosaics of non-overlapping CCD detectors. 
The last column of Table~\ref{obs} gives the number of independent
detectors in each camera when this number is greater than one. 
Therefore, although we have 1,884 images, clearly no individual
star appears in all those images.  In fact, no star appeared in
more than 97 $B$-band images, 240 $V$ images, 51 $R$ images, or 41
$I$ images.  The {\it median\/} number of observations for one of
our stars is 22 in $B$, 72 in $V$, 24 in $R$, and 16 in $I$. 
Since most pointings were centered on or near the galaxy, member
stars typically have more observations than field stars lying
farther from the galaxy center.  

Our complete photometric catalog for \leoi\ and a stacked
image of the field are available at our web site\footnote{
http://www.cadc-ccda.hia-iha.nrc-cnrc.gc.ca/community/STETSON/homogeneous/LeoI.}  

%%%%%%%%%%%%%%%%%%%%%%%%%%%%%%%%%%%%%%%%%%%%%%%%%%%%%%%%%%%%%%%%%%%%
\section{Variable star detection and characterization}\label{variable}

\begin{figure}[!ht]
\begin{center}
\includegraphics[width=0.32\textwidth]{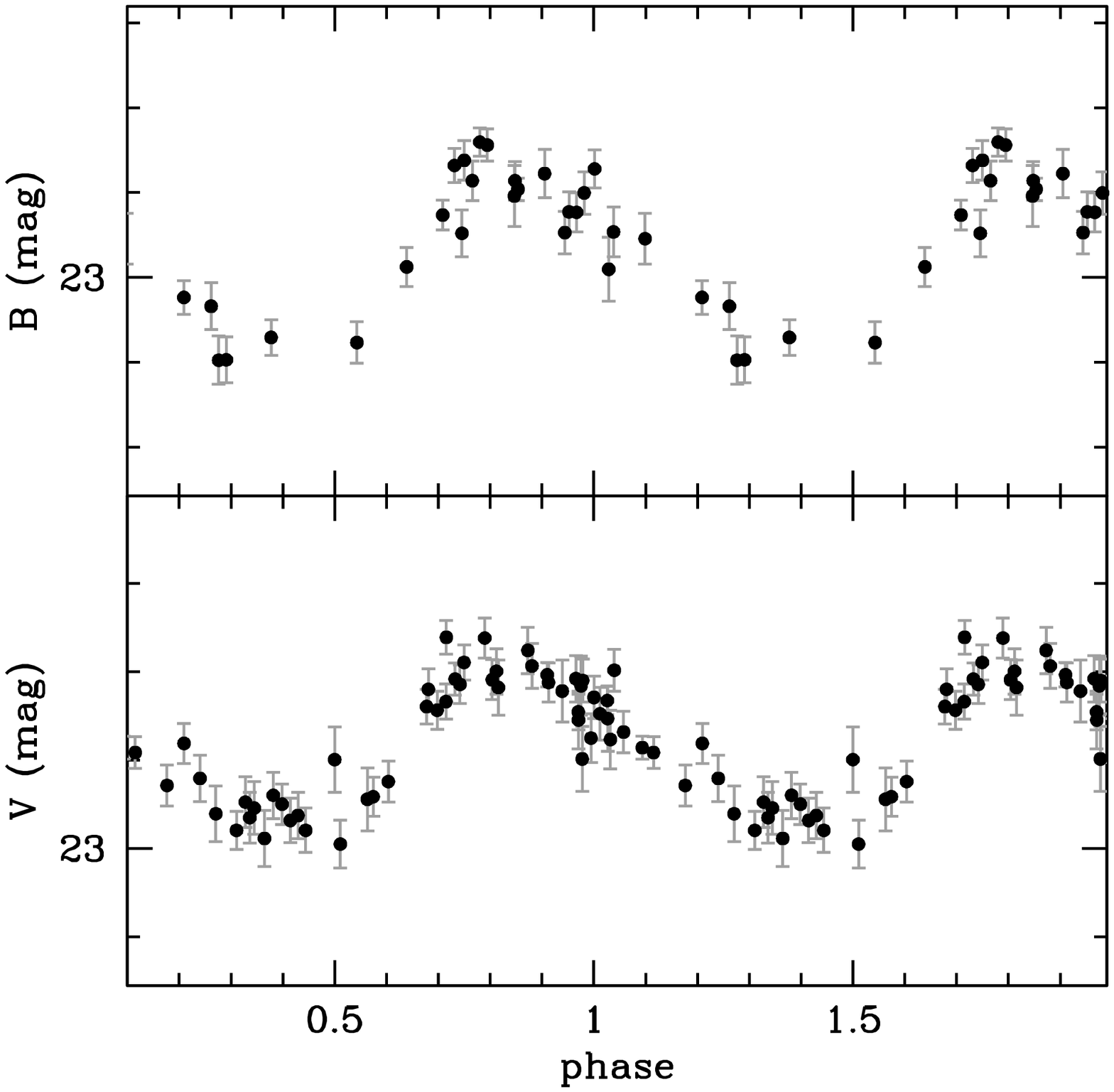}
\includegraphics[width=0.32\textwidth]{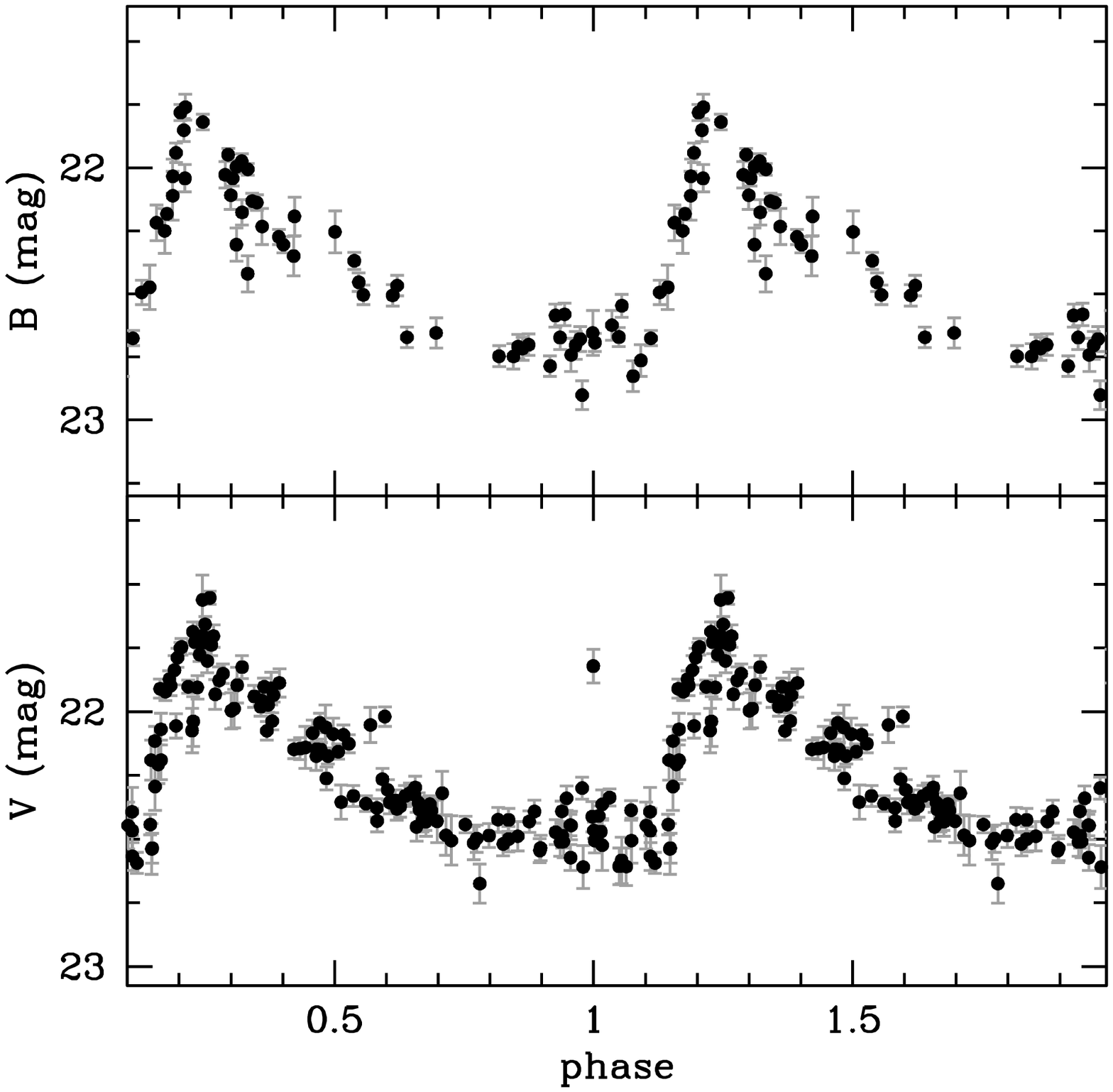}
\includegraphics[width=0.32\textwidth]{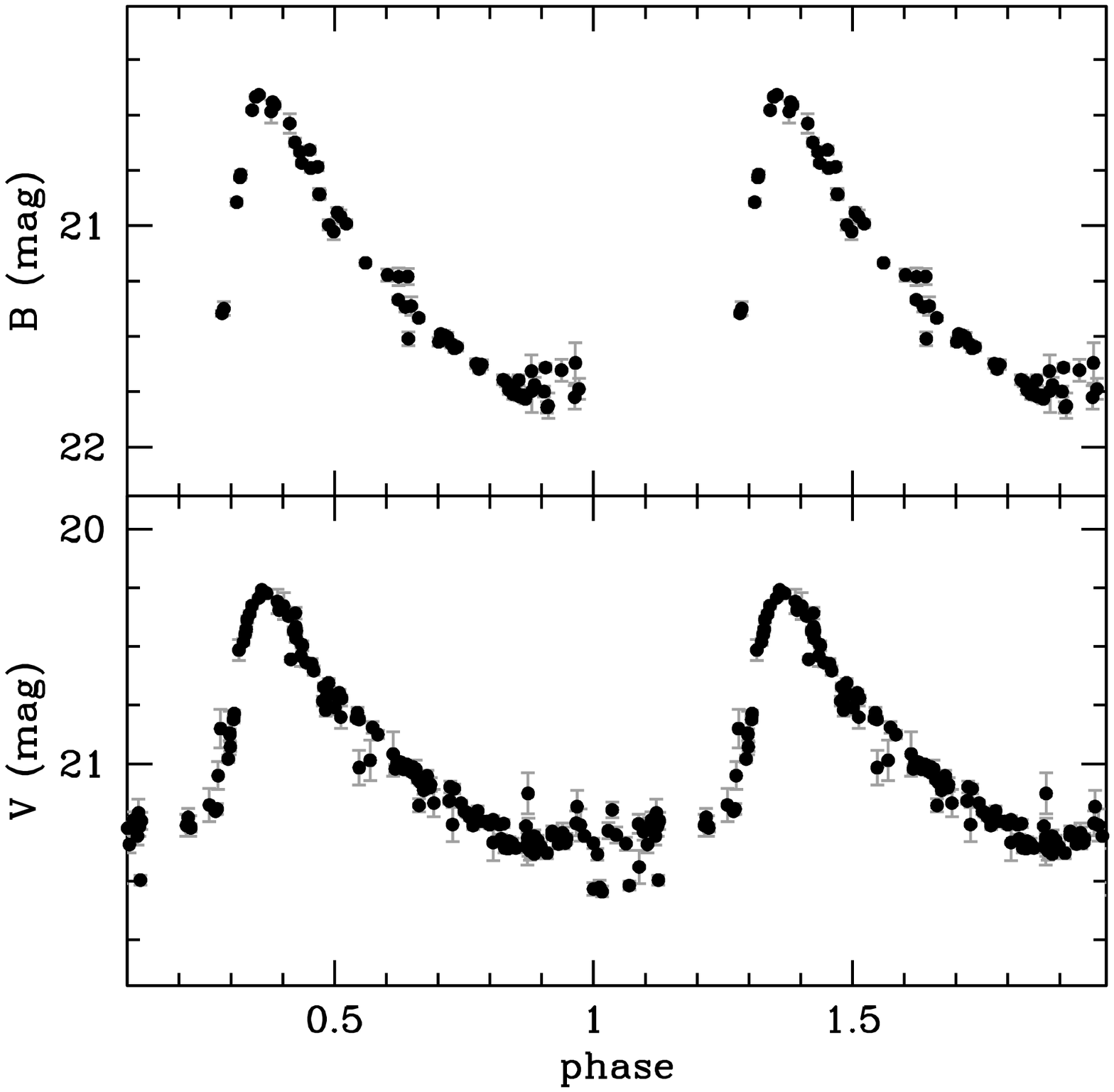}
\caption{Examples of our light curves to illustrate the quality obtained in the $B$
  and $V$ bands. We have selected one RRc (V60, left panel), one RRab
  (V92, possibly a Blazhko or double-mode pulsator candidate, middle) and one AC (V120, right). This last star has the uncommon 1$\,$d
  period, but with our twenty-year baseline aliasing is minimal.  
\label{fig1}
}  
\end{center}
\end{figure}

We have used an updated version of the \citet{welch93} method to
identify candidate variable stars on the basis
of our multi-band photometry of Leo~I. We have identified 251
variable star candidates in a luminosity range that goes from the
Horizontal Branch (HB) to the tip of the red giant branch
(TRGB).  A first guess at the variability period of each star was
found using a simple string--length algorithm \citep{stetson98a},
then a robust least-squares fit of a Fourier series to the data
refined the periods and computed the flux-weighted mean magnitudes
and amplitudes of the light curves \citep[][and references
therein]{stetson98a,fiorentino10a}. The resulting pulsation
properties are listed in Tables~\ref{tabrr},~\ref{tabac} and
~\ref{tablpv}. This list includes 164 stars that we have
classified as RRLs, 55 AC/spCs and 19 LPVs. There are also thirteen
more variable candidates whose physical nature is unclear (Table~\ref{tabcan}).  Some
of their light curves are shown in Fig.~\ref{fig1}, and all are 
available in the on--line version of the paper and at our web site.

On the basis of the photometric quality of the data and the
completeness of the phase coverage we have divided the pulsating
variable candidates into three quality classes, namely: high
confidence or ``good''  (A), uncertain or ``fair'' (B) and very
uncertain or ``poor'' (C).  These grades are given in
Tables~\ref{tabrr} (RR Lyraes) and \ref{tabac} (AC/spCs).  Among
the 164 stars classified as RRL, we have assigned 95 A's, 47 B's,
and 22 C's; among the 55 AC/spCs there are 53 A's, no B's, and 2
C's.  In the case of a star classified as ``A'' we are reasonably
confident that our pulsation properties are ``essentially''
correct; i.e., we believe that the period is correct to better
than $\sim\,$1\%, and the amplitudes are correct to
0.1--0.2$\,$mag. For stars classed ``B'' there is a small
($<\,$25\%?) chance that one or more parameters are wrong by more
than this, and for stars classed ``C'' we feel that the chance of
an error larger than this might be as high as 50\%.  These
quantitative likelihoods are purely the educated guesses of a
fairly experienced observer.  We invite interested readers
to examine the light curves and/or download our original data
from our web site and judge the reliability of our inferences
for themselves.  We anticipate that future reinvestigations based
upon new data will be able to retroactively establish how
meaningful our subjective quality classes truly are.

\begin{figure}[!ht]
\begin{center}
\includegraphics[width=0.70\textwidth]{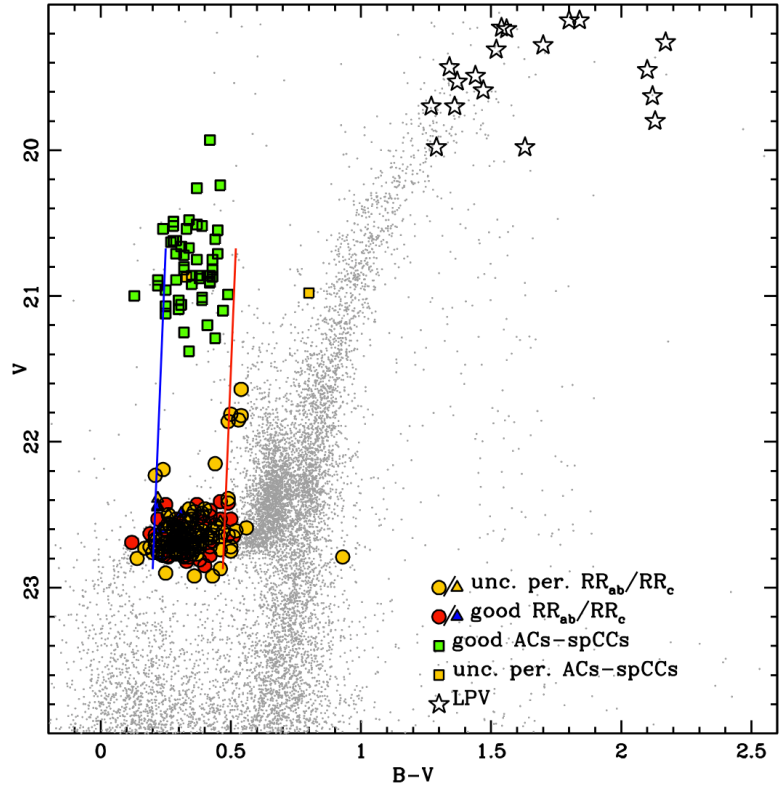}
\caption{CMD and topology of the instability strip for the variable
  stars of Leo~I. Red circles and blue triangles represent RR Lyrae stars
  pulsating in the fundamental and first-overtone mode,
  respectively. Squares and stars represent Cepheids and LPV
  variables. Variable 
  candidates with uncertain periods are indicated by yellow symbols.
\label{fig2}
}  
\end{center}
\end{figure}

In Fig.~\ref{fig2} we show the locations in the CMD of these
variable star candidates and the theoretical instability strip
(IS) derived from radial nonlinear pulsating models for AC/spCs
\citep{marconi04,fiorentino06} and RRLs \citep{dicriscienzo04}. We
have not included here the thirteen variables that we have not
been able to characterize (see Table~\ref{tabcan}).  The agreement
between the locations of ``good'' RRLs (blue and red dots) and
AC/spCs (green squares) and the theoretical boundaries of the IS
is excellent with very few exceptions. We also show the possible
LPV candidates (stars); they are all located near the TRGB and the
upper asymptotic giant branch (see
Table~\ref{tablpv}).  Unfortunately, the sampling of our light
curves for the LPVs is neither dense nor well distributed over the
necessary range of time intervals, and we could not reach any
definitive evaluation of their periods. Detailed periods, being
related to the LPV's masses and ages, could have provided
constraints on the star formation history of the galaxy
\citep[see][and references therein]{feast13}.  The so-called
``periods'' that we give in Table~\ref{tablpv} are formal
maximum-likelihood values returned by our software, but they
should be regarded as purely notional.  We provide them only to
give some sense of the timescales upon which the variations {\it
appear\/} to be occurring.  We commend these stars to the
attention of individuals having extensive access to telescopes of
moderate aperture.

\begin{figure}[!ht]
\begin{center}
\includegraphics[width=0.70\textwidth]{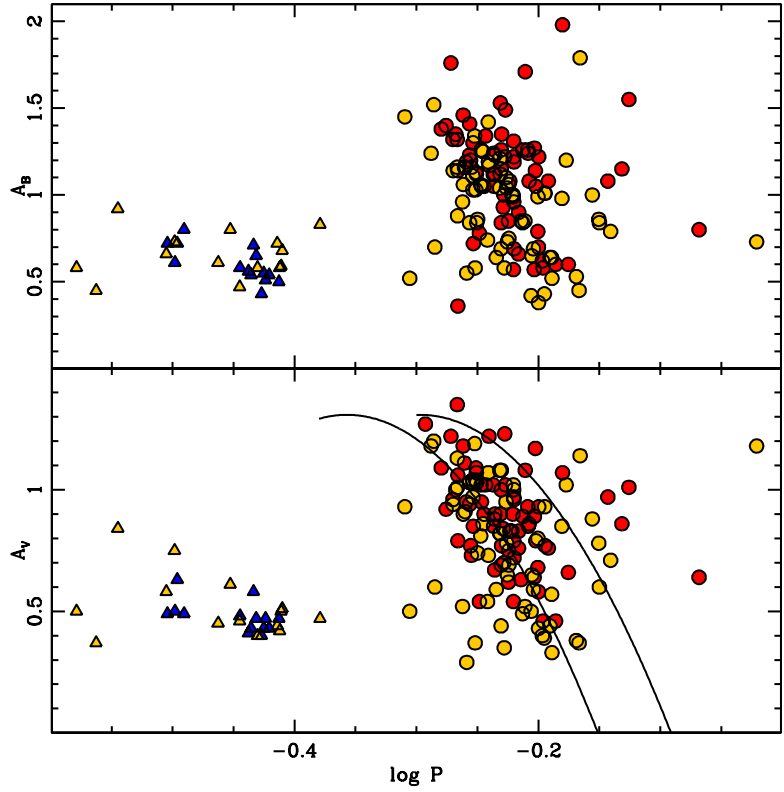}
\caption{Bailey diagram for RRL stars in the \B\ (top panel) and \V\ bands
  (bottom panel). Symbols are as in Fig.~\ref{fig2}. We have also shown,
as solid curves, the
  empirical relations for Oosterhoff types I (left) and II (right) as derived by
  \citet{cacciari05}. 
\label{fig3}
}  
\end{center}
\end{figure}

Our sample of 164 RRL candidates contains 95 good stars (see
Table~\ref{tabrr}) including 24 that apparently either show the
Blazhko effect or pulsate in multiple modes.  Among the good RRLs
we recognize 14 first overtone (FO) and 81 fundamental-mode (FU)
pulsators on the basis of their period/amplitude distribution in
the Bailey diagram (see Fig.~\ref{fig3}). The remaining 69 RRL
candidates with fair and poor light-curve fits are provisionally
divided into 14 FO (12 ``B'' and 2 ``C'') and 55 FU (35 ``B'' and
20 ``C'') pulsators. The pulsation properties of the RRL sample
will be discussed in more detail in the next section.  

The sample of AC/spCs is quite abundant, populating the whole
theoretical IS, and consists of 55 objects (see
Table~\ref{tabac}), including two with uncertain or very uncertain
periods (V212 and V241). Our sample also includes one possible
Blazhko or multiple-mode Anomalous Cepheid, V129. The evolutionary
classification of these stars has been discussed in a companion
paper \citep{fiorentino12d} and will not be repeated here, but in
Section~\ref{cep} we will place some constraints on their masses
and mode classification.  

\begin{figure}[!ht]
\begin{center}
\includegraphics[width=0.70\textwidth]{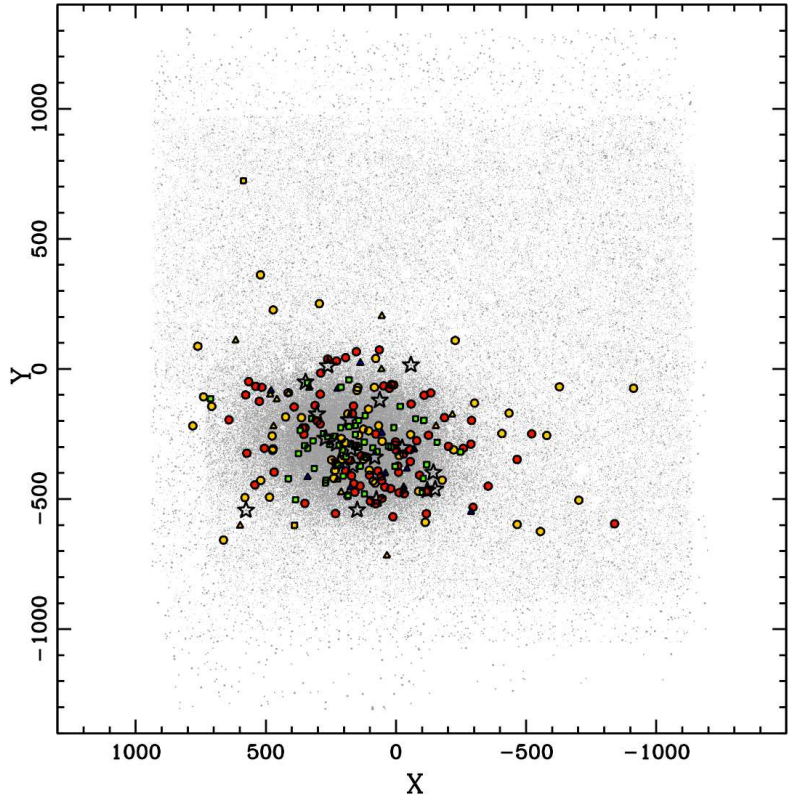}
\caption{Distribution of our variable candidates in
  the field of view of our photometric study. The symbols for variable stars are color-coded
  as in Fig.~\ref{fig2}. Gray dots represent non-variable stars and are expanded
  according to their brightness in the $V$ band.  Units are arcseconds relative to an arbitrary
origin (stated in text); X increases east (to the left) and Y increases north (up). An ellipse represents an isopleth of the stellar distribution with a semimajor axis of 12\min\ as derived by \citet{mateo08}.} 
\label{fig4}
\end{center}
\end{figure}

The spatial locations of all our variables are shown in
Fig.~\ref{fig4}.  The $(x,y)$ coordinates in this diagram are
referred to an origin $(0,0)$ that we have arbitrarily placed at
10\hr08\tm16\Ts21 +12\deg23\min05\Sec0 (close to the center of
an early subset of our images).  The $y$-axis is the great circle
$x = 0$ and is equal to the great circle
$\alpha$~=~10\hr08\tm16\Ts21, and the $x$-axis is tangent to
$\delta$~=~+12\deg23\min05\Sec0 at the origin.  The photocenter
of Leo~I has been estimated to lie at $(\alpha,\delta) =
(152.1125 , 12.30833333)$ (= 10\hr08\tm27\Ts00 +12\deg18\min30\Sec0) according to \citet{mateo98}; we
ourselves estimate the center of \leoi\ to lie at
10\hr08\tm27\Ts13 +12\deg18\min22\Sec2 [$(x,y) = (+160.0,-282.8)$], with an uncertainty of
order 1\sec\ in each coordinate. This estimate has been derived
by determining the median $x$ and $y$ coordinates of probable
member stars lying within a range of distances from the center,
from 100\sec\ to 850\sec; the consistency of these values
provides our estimate of the uncertainty \citep[see][\S3.4]{stetson98b}. Fig.~\ref{fig4} shows
that the RRLs, as representative very old stars, have a larger
spatial extent than the younger, more massive stars that we
presume the AC/spCs and LPVs to be; the latter appear more
centrally concentrated. This supports the idea that AC/spCs in
Leo~I are of the same nature as those in the Large Magellanic
Cloud (LMC), where ACs seem to result from both the intermediate-age stellar population and old
binary systems \citep{fiorentino12c}. This is consistent with our
provisional classification of these stars as a mix of both ACs and
spCs \citep[see][for details]{fiorentino12d}.  

%%%%%%%%%%%%%%%%%%%%%%%%%%%%%%%%%%%%%%%%%%%%%%%%%%%%%%%%%%%%%%%%%%%%

%%%%%%%%%%%%%%%%%%%%%%%%%%%%%%%%%%%%%%%%%%%%%%%%%%%%%%%%%%%%%%%%%%%%
\section{RR Lyrae stars}\label{rr}

The Bailey diagrams are shown in Fig.~\ref{fig3} for the $B$ (top)
and $V$ (bottom) photometric bands. As expected, the RRLs separate
very cleanly into two different groups, the longer-period FU
pulsators (log~P$\,\gsim\,$--0.35, P$\,\gsim\,$0.45$\,$d: RRab or,
to some researchers, RR0) and the shorter-period FO pulsators
(log~P$\,\lsim\,$--0.35: RRc, or RR1).  These diagrams are very
often used to give an indication of the Oosterhoff class. For this
reason, in the $V$-band (bottom) panel we have also plotted the
curves \citep{cacciari05} representative of the Oosterhoff
dichotomy shown by RRab stars in Galactic GCs. We note that
most of the RRab stars seem to follow the OoI line, reaching a
maximum $A_V$ amplitude of about 1.2 mag. However, as discussed
in \citep{fiorentino12d}, based on the mean period of the RRab
sample we have classified Leo~I as Oo-intermediate, since
$\left<\hbox{\rm Pab}\right>\,=$\,0.596$\pm$0.001$\,$d ($\sigma$=0.05), based
on only those stars with good light-curve fits. This result does
not change when we include the RRab variables with fair or poor
periods or when we add the RRc stars with fundamentalized periods
\citep{coppola13}.  

We have also calculated the ratio between the
number of RRc and the total number of RRLs, which is considered
another indicator of Oosterhoff class, ${N_{\rm c}}/{(N_{\rm c} + N_{\rm ab})}
 = 28/164 \approx 0.17$. This value of the ratio includes the variable candidates
with fair and poor light-curve fits; a slightly lower value, $14/95 \approx 0.15$, is obtained
when it is based on only the good RRLs. We have a slight preference
for the higher value, however, because the very classification
``good,'', ``fair,'' and ``poor'' disproportionately
assigns the lower quality classes to the c-type variables with their
smaller amplitudes.  Although those same smaller amplitudes 
probably mean that the c-type variables are more affected by incompleteness
even if we ignore the quality classes, these very small values of 
the c-to-total ratio still suggest that it would be hard for us to have
missed enough c-type variables to push Leo~I out of the the OoI
class.  This conclusion is in line with what has been observed in
other nearby dSph galaxies (see section~\ref{dSph} for details). 

We take advantage of the good sampling of our light curves, in
particular in the $B$ and $V$ bands, to estimate the mean amplitude
ratios $\left< A_V/A_B\right>=0.825\pm0.002\,$mag ($\sigma\,$=$\,$0.20) and
$\left<A_R/A_V\right>=0.763\pm0.004\,$mag ($\sigma\,$=$\,$0.24), based on 95
and 63 RRLs respectively (not all stars with ``A''-quality data in
$B$ and $V$ have good data in $R$). This $A_V/A_B$ value is in full agreement
with the value discussed in \citet{dicriscienzo11} based on 130
RRLs taken from nine GGCs.  

\subsection{Distance to Leo~I from RRLs}\label{dist}

In this section we use the RRL pulsation properties to constrain
the distance to Leo~I.  In particular, we use two independent
methods: a linear formulation of the $M_V$ versus [Fe/H] relation
(\cite{bono03a}; see also \cite{cacciari03a}) and the First Overtone Blue Edge (FOBE) method
\citep[][and references therein]{caputo00}. 

For the $M_V$ versus [Fe/H] relation we consider only the 81 RRab-type 
variables with light curves of quality class ``A'' (``good''), and we
compare their measured properties to the following theoretical
formulations: 
\begin{equation*}
M_V(RR) = (0.18\pm0.07)\hbox{\rm [Fe/H]}+(0.72\pm0.07),\quad \hbox{\rm for [Fe/H]$\,\leq\,-1.6\,$dex}
\end{equation*}
and 
\begin{equation*}
M_V(RR) =(0.36\pm0.03)\hbox{\rm [Fe/H]}+(1.04\pm0.08),\quad \hbox{\rm for [Fe/H] $>$ --1.6 dex}
\end{equation*}
\citep{bono03a}.
In order to use these formulae we must assume a metal abundance
for Leo~I. Recent measurements of the galaxy's metallicity based
on a large ($\sim\,$850) sample of individual RGB stars studied
with medium-resolution spectroscopy have provided a mean
metallicity of [Fe/H] $\sim$ --1.43 \citep[$\sigma =$ 0.33
dex;][]{kirby11}. However, it is worth mentioning that Leo~I seems
to have a broad metallicity distribution, with iron abundances
ranging from $\sim\,$--2.15 to $\sim\,$--1. RRLs being among the
oldest and presumably the metal-poorest stars in a stellar system,
we decided to derive two different distance moduli corresponding
to both the average Leo~I metallicity and the metal poor extreme.
We found $\left<V\right> = 22.648 \pm 0.001\,$mag ($\sigma=0.11$)
for the RRL sample, and assuming E(\bmv) = 0.02 mag\footnote{
The adopted value is taken from \citet{burstein84} and is equal
to a mean of different estimates used in the literature
\citep{gallart99,bellazzini04,held01}. We note that the use of the
recent and slightly higher reddening value provided by
\citet{schlafly11} (0.031$\,$mag) would cause a decrease of
$\sim$0.03 mag in the true distance modulus estimated here.}, the
two distance moduli turn out to be $\mu_0=22.06\pm0.08$ and
$22.25\pm0.07\,$mag for [Fe/H]$\,=\,$--1.43 and --2.15,
respectively. These values are in good agreement with the distance
modulus derived by \citet{bellazzini04} using the TRGB
method\footnote{In this context it is worth noting that the
TRGB method has a minimal dependence on metallicity in the
relevant range, $-2.0 < \hbox{\rm [Fe/H]} < -1.2$; \citep[see, for
example,][]{tammann08}.}: $\mu_0=$22.02$\pm$0.13 mag,
corresponding to a heliocentric distance
D$\,=\,254^{+16}_{-19}\,$kpc.  

\begin{figure}[!ht]
\begin{center}
\includegraphics[width=0.70\textwidth]{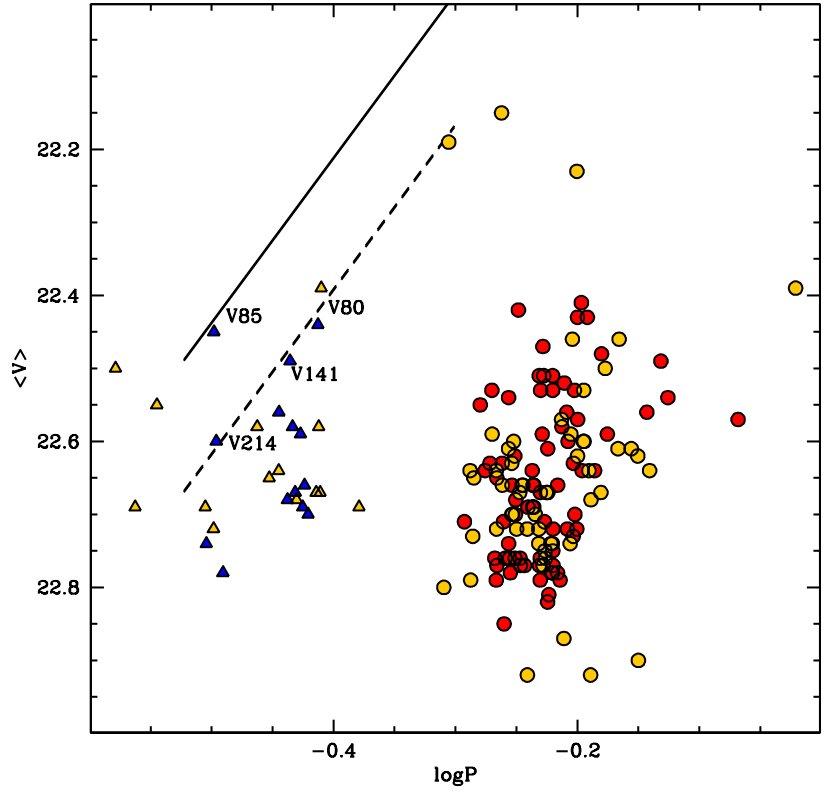}
\caption{The period-luminosity distribution of RRLs in \leoi. The symbols are
  color-coded as in Fig.~\ref{fig2}. The solid line corresponds to
  the FOBE position as constrained by V85, whereas the dashed one is
  drawn to match V80, V141, and V214.
\label{fig5}
}  
\end{center}
\end{figure}

Now we use the technique extensively discussed in
\citet{caputo00}, which is a graphical method based on the
predicted period-luminosity (PL) relation for pulsators located
along the FOBE.  It seems quite robust for clusters with
significant numbers of RRc variables and is thus applicable to
\leoi.  The distance modulus is derived by matching the observed
distribution of RRc variables to the following theoretical
relation: 
\begin{equation*}
M_V\hbox{\rm\footnotesize (FOBE)} =-0.685(\pm0.027) - 2.255 \log P\hbox{\rm\footnotesize (FOBE)}
 - 1.259 \log (M/M_{\odot}) + 0.058 \log Z.
\end{equation*}
Here, we assume the same two metallicity values for the RRLs in
Leo~I and adopt M = 0.7M$_{\odot}$ from the evolutionary HB models
for RRc variables, with an uncertainty of the order of 4\%
\citep{bono03a}.  The comparison between the observed RRLs and the
theoretical relation is shown in Fig.~\ref{fig5}. Given the
period-luminosity distribution of our RRc sample and the high
uncertainty in the FOBE determination, we decided to use two
possible FOBE evaluations, the first defined by the ``good'' RRc
V85 and the other by V80, V141 and V214. Thus, the FOBE method
returns a range of possible distance moduli, $\mu_0=$21.96 (V85)
to 22.14 (V80,V141,V214) assuming [Fe/H]$=$--2.15 and E(\bmv) =
0.02 mag; these decrease by 0.04 mag when using the metal-rich
stellar content (--1.43 dex). It is worth noting that the
distance evaluation based on the FOBE defined by V85 gives a
distance determination that is also in very good agreement with
that given by the TRGB method \citep{bellazzini04}.

%%%%%%%%%%%%%%%%%%%%%%%%%%%%%%%%%%%%%%%%%%%%%%%%%%%%%%%%%%%%%%%%%%%%

%%%%%%%%%%%%%%%%%%%%%%%%%%%%%%%%%%%%%%%%%%%%%%%%%%%%%%%%%%%%%%%%%%%%

\section{Anomalous and short-period Cepheids}\label{cep}

In a previous paper \citep[][]{fiorentino12d}, we discussed
the evolutionary classification of the Cepheid sample we have
detected in Leo~I. On the basis of a comparison with a set of
evolutionary tracks at different metallicities ([Fe/H] ranging
from --1.8 to --1.0) we have concluded that we are dealing with an
unprecedented mix of ACs and spCs.  In this section, we discuss
the pulsation properties of this unique sample of variable
stars.\par 

\begin{figure}[!ht]
\begin{center}
\includegraphics[width=0.70\textwidth]{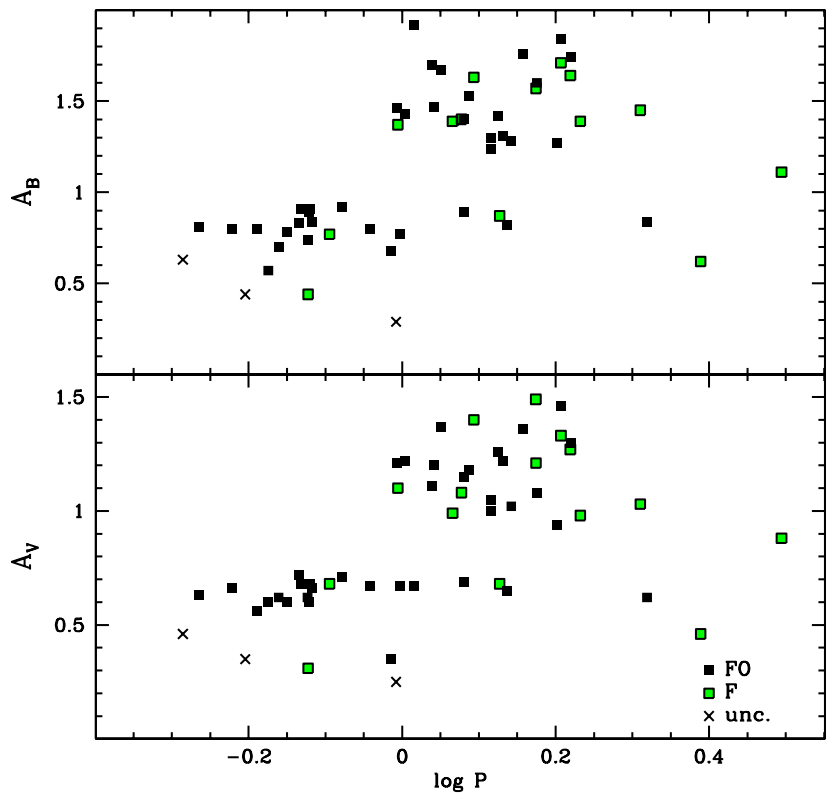}
\caption{The same as in Fig.~\ref{fig3} but for Cepheids. Both amplitude
  distributions, $A_V$ (bottom panel) and $A_B$ (top panel),
  clearly show that
  our Cepheid sample forms two clusters of stars distinguished by 
  small amplitudes for log$\,$P$\,\lsim\,$0, and larger amplitudes for 
  log$\,$P$\,\gsim\,$0. Stars are color-coded according to their classification as
  derived in Section~\ref{cep}. Filled green and black squares correspond to
  FU and FO pulsators respectively. Crosses indicate those stars with uncertain mode
  classification (see the text for details).  
\label{fig6}
}  
\end{center}
\end{figure}

In Fig.~\ref{fig6} we show their distribution in the amplitude versus
period diagrams in the $B$ (top) and $V$ bands (bottom). It seems
clear that the sample can be divided into high- and low-amplitude
subsamples separated at $A_B\sim1.1$ and $A_V\sim0.8$, with the
low-amplitude stars dominating for periods less than one day and
the high-amplitude stars dominating at longer periods.
Similar behavior seen in
other classes of pulsating variables, e.g., RRLs, usually identifies a mode
separation between the first overtone (lower amplitudes, shorter periods) 
and fundamental (higher amplitudes, longer periods)
modes of pulsation. However, for ACs the mode classification is far
from trivial and cannot be easily linked to either the period-amplitude
distribution or the morphology of the light curves, as
discussed in \citet{marconi04}. 

We note that observational uncertainties probably do not 
contribute much to the scatter of points in the amplitude versus
period diagrams.  For the 53 out of 55 candidates to
which we have assigned quality class ``A,'' we feel that the
amplitude uncertainties are probably not worse than 0.1$\,$mag and
the period uncertainties are probably not larger than 1\% ($\sigma(\log P) \sim 0.004$).
These numbers are much smaller than the dispersion seen in Fig.~\ref{fig6}.
We conclude, therefore, that the spread is real and intrinsic to the
stars themselves.

\begin{figure}[!ht]
\begin{center}
\includegraphics[width=0.70\textwidth]{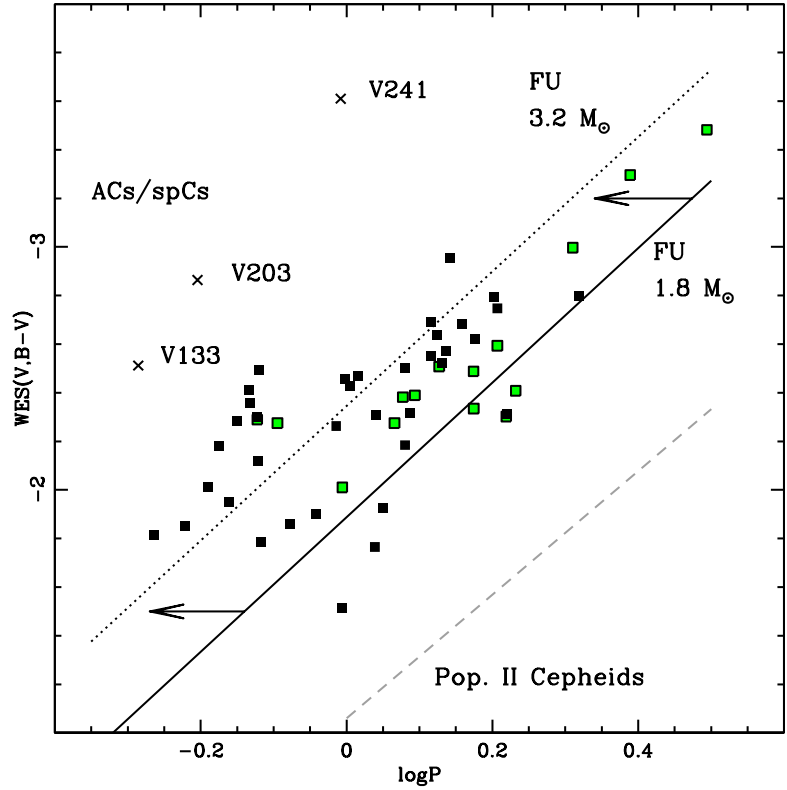}
\caption{The Wesenheit distribution for RRLs and
  Cepheids. The same color code as in Fig.~\ref{fig6} has been used.
The theoretical Wesenheit relations for stars with masses of 1.8 and 3.2~M$_\odot$ pulsating in the fundamental mode 
  (solid and dotted lines, respectively) are shown, as
  predicted by \citet{fiorentino06}. Arrows indicate the offset of the 1.8~M$_\odot$ first-overtone
(FO) locus from the FU locus.  For comparison, the location of
  Population II Cepheids is also indicated by a dashed line \citep{dicriscienzo07}. 
\label{fig7}
}  
\end{center}
\end{figure}

The distribution in the period-luminosity diagram might help in
the mode classification. For this reason, in Fig.~\ref{fig7} we
show the AC/spC distribution in the reddening-free ($V$,\bmv)
Wesenheit magnitude versus period plane. We have chosen this color
combination because the available $B$ and $V$ observations best sample
the light curves, resulting in the most accurate mean magnitudes.  
Uncertainties of a few$\,\times\,$0.01$\,$mag in the adopted
mean $B$ and $V$ magnitudes for the stars lead to uncertainties
$\lsim\,$0.1$\,$mag in the derived Wesenheit magnitudes.
For comparison we have also shown the Wesenheit relations
predicted by theoretical models for ACs assuming masses of 1.8 and
3.2 M$_\odot$ \citep{marconi04,fiorentino06} and metallicities Z
from 0.0001 to 0.0004, as well as those for Population II Cepheids
\citep{dicriscienzo04}. We have assumed $\mu_0 =$22.11 $\pm$ 0.15
with a reddening of E(\bmv)=0.02 mag, values that were used by
\cite{fiorentino12d} to classify our sample of variable stars and
that agree quite well with what is found using RRLs (see previous
section). Our sample occupies the general region where ACs are
expected without defining any tight sequence in this diagram.
There are also no distinct subdistributions forming different
sequences of AC/spCs such as are typically observed in other
galaxies, where abundant samples of ACs \citep{fiorentino12c}
and/or spCs \citep{bernard13} show that both FU and FO pulsators
are detected.  

\begin{figure}[!ht]
\begin{center}
\includegraphics[width=0.70\textwidth]{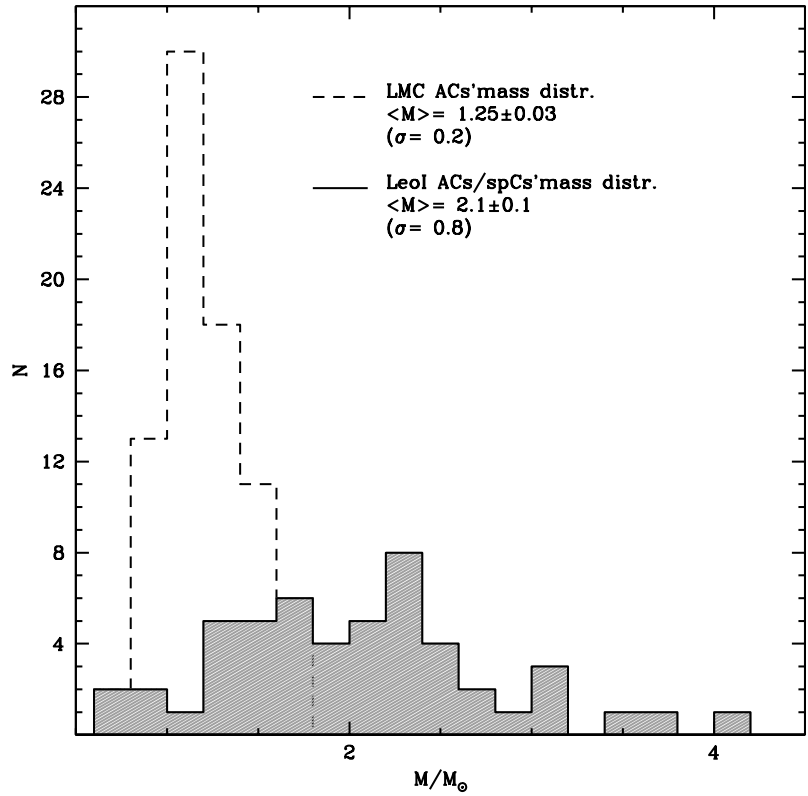}
\caption{The mass distribution for \leoi~ Anomalous and
short-period Cepheids (black solid lines and gray fill) has been computed using
the approach discussed in \citep{fiorentino12c}. The mean mass and the standard deviation
reported in the legend exclude six outliers with derived masses larger
than 5 M$_\odot$. For comparison, the dashed lines show the mass
distribution of ACs in the OGLE sample for the LMC. The mean mass
of the LMC Cepheids has also been specified in the legend.
\label{fig8}
}  
\end{center}
\end{figure}

To investigate this behavior, we again use
the approach discussed in \citet{fiorentino12c}, one that
allows us to identify the mode classification within the sample of
ACs in the LMC released by OGLE III \citep{soszynski08c}. This
method, detailed in \citet{caputo04}, returns simultaneously the
mode and the pulsation mass for each individual AC. It is based on
the use of a theoretically predicted mass- and luminosity-dependent relation
between period and $V$-band amplitude (MPLA) for FU mode pulsators,
which is {\it not\/} followed by pulsators in higher modes.
Coupling this relation with the mass-dependent
period-luminosity-color (MPLC) relation that exists for both
modes, we will assign the FU pulsation mode only when the two
masses agree within 1$\sigma$. We use the following relations: 
$$\log P_F + 0.41 M_V + 0.77 \log\hbox{\rm M} = 0.01 - 0.188A_V,$$
\noindent predicts the visual amplitude from the period, luminosity and mass for fundamental pulsators, and
$$M_V = -1.56 - 2.85\log P + 3.51(\bmv) - 1.88 \log\hbox{\rm M}\quad\hbox{\rm (FU)},$$
$$M_V = -1.92 - 2.90\log P + 3.43(\bmv) - 1.82 \log\hbox{\rm M}\quad\hbox{\rm (FO)}$$
\noindent predict the luminosity from the period, color and mass for FU and FO pulsators.
These have been derived in \citep{marconi04}, and can be inverted to yield
$$\log[\hbox{\rm M/M$_\odot$\footnotesize{(MPLA-FU)}}]=\bigl(0.01-0.188 A_V-\log P-0.41 M_V\bigr)/0.77$$
$$\log[\hbox{\rm M/M$_\odot$\footnotesize{(MPLC-FU)}}]=-\bigl(M_V+1.56+2.85\log P-3.51(\bmv)\bigr)/1.88$$
$$\log[\hbox{\rm M/M$_\odot$\footnotesize{(MPLC-FO)}}]=-\bigl(M_V+1.92+2.90\log P-3.43(\bmv)\bigr)/1.82.$$
Our application
of these equations produces the masses and the classifications given in
Table~\ref{tabac} and summarized in Fig.~\ref{fig8}. In this
figure we show the mass distribution (gray histogram) predicted
for our sample based on the pulsation relations. This is to be
compared with the very peaked mass distribution shown by ACs
observed in the LMC. This confirms our interpretation that a
continuous spread in masses in \leoi\ causes the dispersion shown in the
Wesenheit plane (see Fig.~\ref{fig7}), whereas in the LMC you can
perceive two distinct sequences in the same plane
\citep{soszynski08c,fiorentino12c}. 

Three AC/spCs in Leo~I show unreasonably high masses, larger
than 5 M$_{\odot}$ (indicated with crosses in Figs.~\ref{fig6} and
~\ref{fig7}); in Fig.~\ref{fig7} these stars are the ones
deviating the most from the global distribution in the Wesenheit
plane due to their very short periods compared to other stars with
comparable luminosities and colors. In particular, the star with
the highest indicated mass, V241, belongs to our ``very
uncertain'' class (C) and shows a very red color
(\bmv$\,\ge\,$0.6), thus suggesting the possibility of problems in
our magnitude, color and/or period determinations for this star.
The other two stars, namely V133 and V203, show very small
amplitudes (see Fig.~\ref{fig6}) suggesting that they might be
slightly blended.  A visual examination of these stars in our
stacked image for \leoi\ suggests that a blending explanation for
their anomalous masses is plausible, but not definite.

%%%%%%%%%%%%%%%%%%%%%%%%%%%%%%%%%%%%%%%%%%%%%%%%%%%%%%%%%%%%%%%%%%%%

%%%%%%%%%%%%%%%%%%%%%%%%%%%%%%%%%%%%%%%%%%%%%%%%%%%%%%%%%%%%%%%%%%%%
\section{Comparing Leo~I with classical spheroidal and ultra-faint
  dwarf galaxies}\label{dSph}
\begin{figure}[!ht]
\begin{center}
\includegraphics[width=1.0\textwidth]{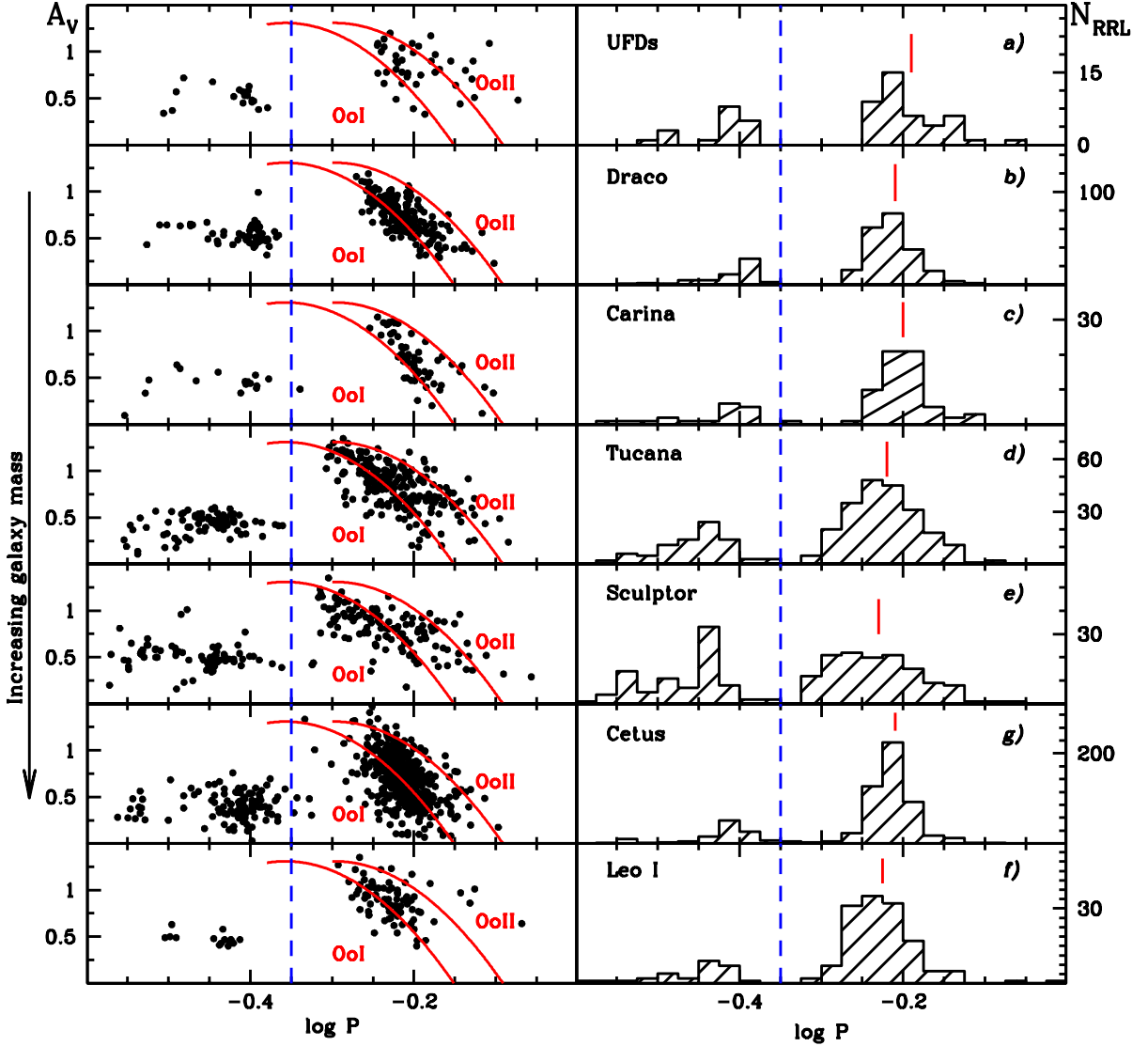}
\caption{{\it left --} Period versus $V$-band amplitude diagram for RRLs
  observed in eleven UFDs (top panel) and six dSphs (remaining six panels) for which mostly complete samples of 
variable stars have recently been published. The galaxies have been ordered by increasing baryonic mass from
top to bottom.  The UFDs have been considered 
  together to improve the statistical significance of the plot. {\it right
    --} Period-frequency distributions of the same RRLs; the mean period of the RRab variables is indicated
  by a vertical red line in each panel.  The vertical blue dashed lines in this
plot indicate $\log P = -0.35$, which we have taken as the nominal boundary between RRc (FO) and RRab (FU)
variables.
\label{fig9}
}  
\end{center}
\end{figure}

\begin{figure}[!ht]
\begin{center}
\includegraphics[width=1.0\textwidth]{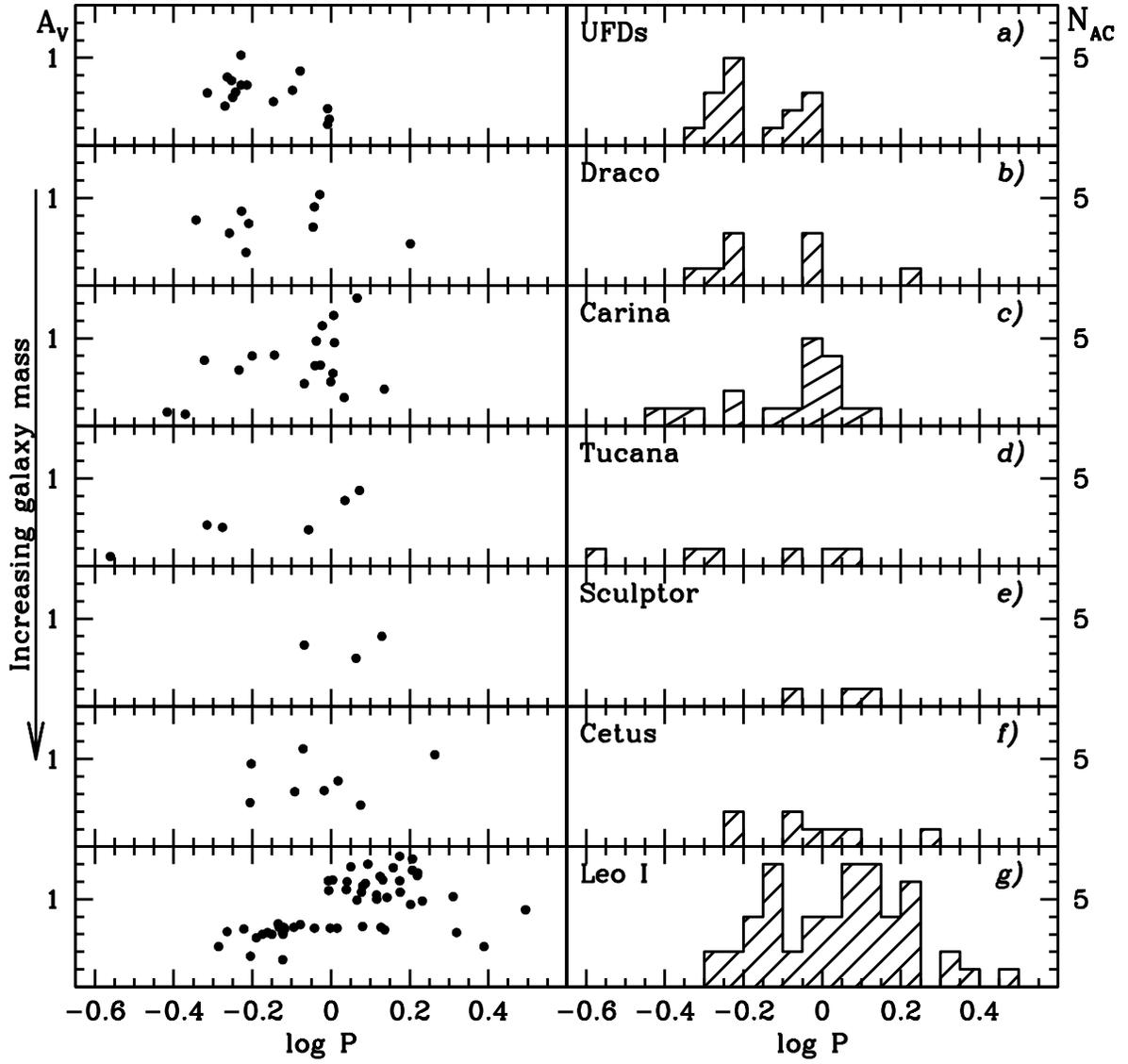}
\caption{Same as Fig.~\ref{fig9} for the AC/spC sample.  
\label{fig10}
}  
\end{center}
\end{figure}

%%%%%%%%%%%%%%%%%%%%%%%%%%%%%%%%%%%%%%%%%%%%%%%%%%%%%%%%%%%%%%%%%%%%
In Figs.~\ref{fig9} and~\ref{fig10} we show the amplitude versus period diagrams
(left) and the period histograms (right) for RRL and AC/spC stars in six
dSphs ordered by increasing baryonic mass \citep{mcconnachie12}, namely Draco
\citep{kinemuchi08}, Carina \citep{coppola13}, Tucana
\citep{bernard09}, Sculptor \citep{kaluzny95}, Cetus
\citep{bernard09,monelli12b} and Leo~I (this paper). In the top panel of each
figure we have also shown all the RRLs and AC/spCs observed in
the eleven UFDs surveyed for variability
\citep{dallora06,kuehn08,moretti09,musella09,dallora12,musella12,clementini12,garofalo13,boettcher13}.  
For comparison, in Fig.~\ref{fig9} we also show the two curves that
trace the Oosterhoff dichotomy observed in Galactic GCs as defined by
\citet{cacciari05}. Lying in general between the two Oosterhoff
curves, the ab-type RRL distributions for the classical dwarfs in the
Bailey diagram are very similar to each other. The only (slight) exceptions
are Cetus and Carina, which show a different slope in the period versus
amplitude plane as discussed in \cite{bernard09} and \cite{monelli12b}. 
For most dSphs, the
period distributions of RRab stars (right panels of Fig.~\ref{fig9}) are quite peaked
around their mean value with two exceptions, viz.\ Sculptor and
Tucana. In Table~\ref{tabmean} we have listed the mean
properties of the RRLs according to the catalogs we have adopted;
the mean RRab and RRc periods are the same within 1$\sigma$.
In the last column we have also listed the total masses and mean
metallicities of these galaxies according to the references used
in \citet{mcconnachie12}. The mean periods of RRLs in UFD galaxies
seem to generally follow the same behavior as in classical dwarfs,
occupying the same general location in the Bailey diagram with a
slightly higher mean period for the RRab stars, very similar to
that of Carina \citep{dallora03,coppola13}. This suggests that,
at least in terms of their RRL properties, there is
not a significant difference between classical and ultra-faint dwarf galaxies.  

On the basis of their similar general properties, we decided to
build up a single large sample of well studied RRLs in those dwarf
galaxies where the variability surveys do not seem to suffer from
strong completeness problems, thus very likely describing the RRL
properties of these stellar systems rather well. This initial
sample contains 1,726 objects (with 1,299 ab-type variables); we
included Cetus and Carina because their inclusion does not bias
the mean properties of the sample nor change our final conclusions.  

As discussed in \citet{fiorentino12d}, different conclusions
result from an inspection of Fig.~\ref{fig10}, where it is clear
that the AC/spCs have different period distributions and Bailey
diagrams. In particular, we note that the period distribution
seems to move toward longer periods when the baryonic total mass
of the galaxy increases (see Table~\ref{tabmean}). This is easily
understood when one considers that the high masses typical of ACs
\citep[1.2$\lsim$M/M$_{\odot} \lsim$ 2.1]{fiorentino12d} may
result from either or both the interactions of old binary stellar
systems and single-star evolution in purely intermediate-age
populations.  Thus, their specific frequency could depend on the
total mass of the host galaxy and on the relative importance of
star formation events at intermediate ages ($\lsim$ 5$\,$Gyr), as
extensively discussed by \citet{fiorentino12c}.  

%%%%%%%%%%%%%%%%%%%%%%%%%%%%%%%%%%%%%%%%%%%%%%%%%%%%%%%%%%%%%%%%%%%%
\section{Comparing dwarf galaxies with globular clusters and the halo field}\label{GCfield}
%%%%%%%%%%%%%%%%%%%%%%%%%%%%%%%%%%%%%%%%%%%%%%%%%%%%%%%%%%%%%%%%%%%%

Usually, the average properties of RRLs in individual dSphs and
UFDs---such as the mean periods of both RRab and RRc stars---are
compared to those observed in GCs as representative of the
Galactic halo \citep[][and references
therein]{catelan09,clementini10}. However, given their different
total stellar masses, GCs typically host RRL populations of order a
factor 10 smaller than those observed in dwarf galaxies, 
making a proper statistical comparison frequently difficult.
Moreover, in the last ten years or so, GCs have been demonstrated
to be quite complex stellar systems despite what we previously
believed in terms of their chemical enrichment histories
\citep{gratton04,piotto05,milone12,monelli13}, and they
may not fairly represent the properties of halo field stars.  Even
though the net effect of their complex histories on their global average
chemical abundances (and therefore on the pulsation properties of
their RRLs) may be negligible, this has not yet been demonstrated. 
Finally, although it seems trivial that the statistical meaning of
averaged properties is, by definition, more significant when a large
sample of objects is considered, we must remember that small
number statistics become quite relevant in those galaxies with
very few {\it confirmed\/} member RRLs, as is the case for most UFDs
\citep{moretti09,musella09,dallora12,clementini12,boettcher13}
where even the term ``average'' approaches meaninglessness.

The justification for our assembling a large sample of RRLs in
classical and ultra-faint dSphs is driven in large part by the opportunity
to compare {\it directly\/}, for the first time, their period and
amplitude distributions with those of a huge RRL sample
representing the Galactic halo.  RRLs are ancient stars, older
than 10~Gyr.  Their younger selves were born during the early
millennia of the Universe, and their testimony can provide
important information about chemical evolution during the early
stages of the assembly of the Galactic halo.  Moreover, they are
their own robust distance indicators that might also reveal
details of the halo's spatial structure
\citep{layden94,kinemuchi06,drake13,zinn14}.\par 

With this goal in mind, we have collected several catalogs---mainly on the
basis of the availability of robust periods and reliable $V$-band amplitudes---that,
taken together, provide about 14,000 RRab stars spanning distances
from $\sim\,$5 to $\sim\,$80 kpc, namely the QUEST
\citep{vivas04,zinn14}, NSVS \citep{wozniak04}, ASAS
\citep{szczygiel09} and CATALINA surveys \citep{drake13}.  In
cases where a star appeared in multiple surveys, we have
retained only those data originating in the most recent and
complete study available.  

We have computed the three-dimensional Galactocentric position of
each individual RRL, first converting (RA,Dec) to (l,b)
coordinates and then assuming for the stars in the very inner part
of the Halo (d$_G\,\lsim\,$7.5 kpc) a
mean metallicity of [Fe/H]$\,=\,$--1.3 dex, while for the external part
we used --1.6 dex, according to the metallicity gradient reported
by \citet{layden94}. To account for the individual interstellar
extinctions we have used the \citet{schlegel98} maps, following
the prescription given in their website. Finally we have applied
the same magnitude versus metallicity relation for the RRLs as we used in
Section~\ref{rr} for [Fe/H]$\,>\,$--1.6 dex \citep{bono03a}, assuming a distance
to the Galactic Center of 7.94$\,$kpc \citep{eisenhauer03,groenewegen08,matsunaga11}.  

In our final catalog, we have not included the samples from \citet{miceli08} and \citet{sesar13}
because amplitudes on their filterless magnitude system must be transformed into $V$ (Landolt)
amplitudes using a metallicity- and temperature-dependent scale factor that could affect the stars' apparent
distribution in the Bailey diagram.  On the other hand, we do include the CATALINA
sample, while acknowledging that it has a bias in amplitudes; we retain only RRLs with
amplitudes $A_V$ larger than 0.4 mag. However, this is the
largest, deepest, and most homogeneous catalog at our disposal, and
its known bias turns out to have a negligible effect on the following
discussion. Finally, we use only ab-type RRLs because they
are less affected by both time-sampling and completeness
problems.  Although not 100\% complete, this final huge catalog
will allow us to make the most comprehensive analysis of
the Galactic halo using field RRLs possible so far. 

In order to draw the most complete global view of the globular
clusters belonging to the Galactic halo, we have decided to
exclude all the RRLs listed in the updated catalog of
\citet{clement01} that are observed in GCs belonging to the
Galactic bulge, i.e., those having both Galactocentric distance
d$_G\,\le5\,$kpc and distance from the Galactic plane Z$\,\le\,$1.5
kpc.  Two exceptions to this rule are M62 (\ngc{6266}) and M28
(\ngc{6626}) that---even if projected onto the bulge region
\citep{harris96}---can not be considered representative of the
bulge population \citep[][and reference
therein]{casetti-dinescu13}.  We have also excluded \ngc{2419}
because it is located in a region of the halo not well covered by
the field-star surveys considered here. Finally, we have included
only GCs for which CCD photometry in $V$, including amplitudes, is
available. The full sample consists of 1,617 RRLs (1,054
ab-type) residing in 35 GCs (see Table~\ref{tabcomp}).  

\begin{figure}[!ht]
\begin{center}
\includegraphics[width=1.0\textwidth]{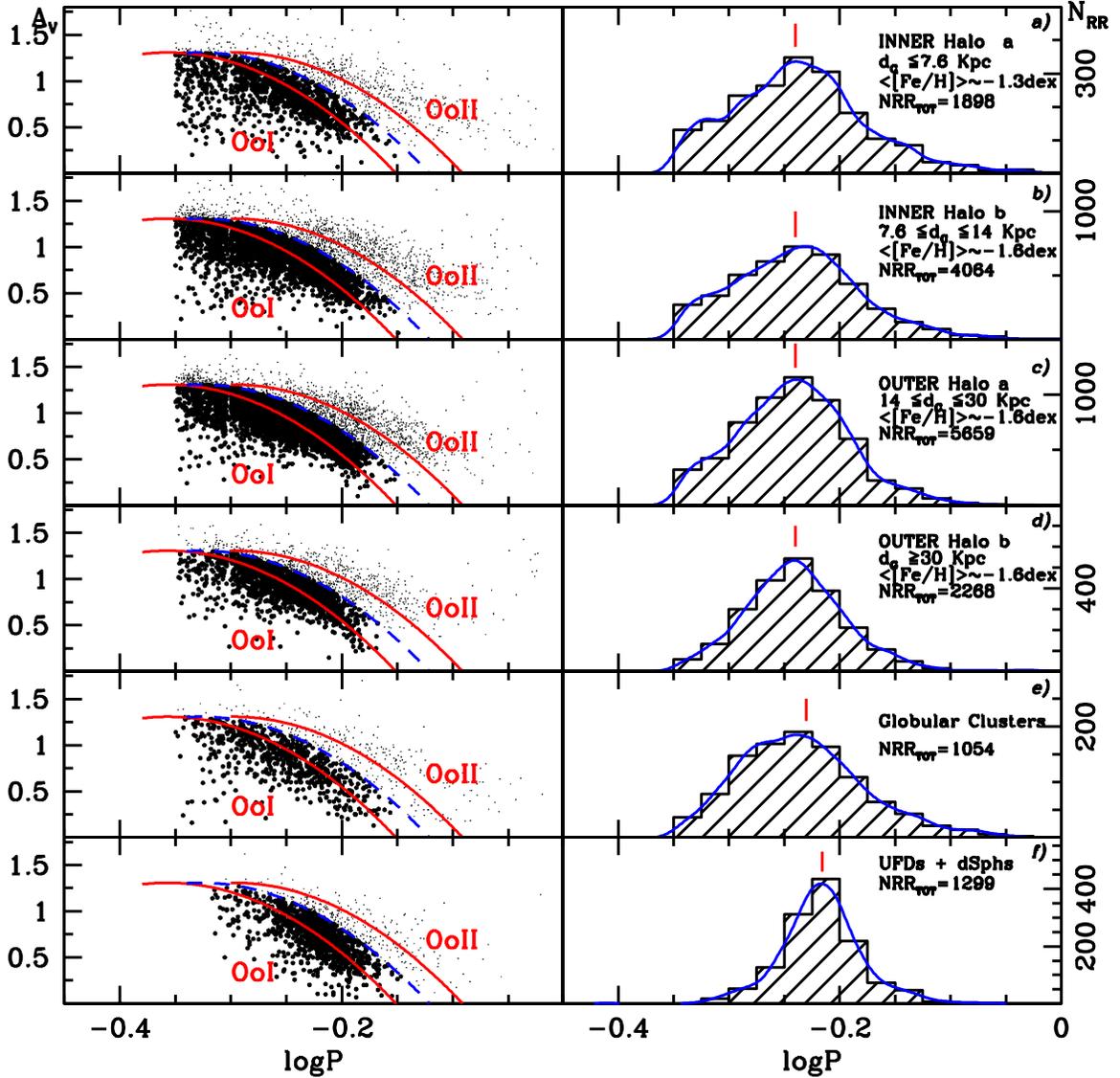}
\caption{{\it left --} Period versus $V$-band amplitude diagram for RRab
  observed in the Galactic halo (panels a, b, c, d)), compared to GCs
  (panel e) and dSphs plus UFDs (panel f). The two Oosterhoff curves
  given in \citet{cacciari05} are also indicated together with a mean
  curve used to notionally separate the OoI from the OoII components (blue dashed line). {\it right
    --} Period distributions of RRab for the same samples plotted on
  the left. In each panel the smoothed histogram used to
  perform the statistical comparison discussed in the text are also plotted (blue solid curves).  
\label{fig11}  
}
\end{center}
\end{figure}

In Fig.~\ref{fig11} we show the distributions of RRab stars in
period-amplitude diagrams (left) and period histograms (right) for
the full set of catalogs we have collected. The first four panels
are devoted to the Galactic halo divided into four
non-overlapping regions: 1) the inner halo a, consisting of stars
with d$_G\,\lsim\,$7.5 kpc (panel a); 2) the inner halo b, stars
with 7.5$\,\lsim\,$d$_G\,\lsim\,$14 kpc (panel b); 3) the outer
halo a, stars with 14$\,\lsim\,$d$_G\,\lsim\,$30$\,$kpc (panel c);
4) the outer halo b, stars with d$_G\,\gsim$30$\,$kpc (panel d).
The boundary between inner and outer halo (d$_G\,\sim\,$14$\,$kpc)
has been chosen according to \citet{kinman12}, whereas the other
reference distances are arbitrary. In the last two panels, we have
plotted the full distributions obtained for GCs (panel e) and
UFDs+dSphs (panel f).  As in Fig.~\ref{fig9}, we have overplotted
the Oosterhoff curves given by \citet{cacciari05}. The average of
their two curves (blue dashed curve in Fig.~\ref{fig11}) allows us
to roughly separate the two populations of OoI (large dots) and
OoII (small dots) RRLs. We have computed the ratio between OoI and
the total number of RRLs, as given in the last column of
Table~\ref{tabcomp}.  These numbers are very similar to each other
and they are compatible with a global OoI classification, in
agreement with recent and independent results from the ROTSE
\citep{miceli08} and LINEAR \citep{sesar13} surveys.  

We have also used these data to compute the mean periods of the
RRLs with \hbox{log$\,$P$\,\ge\,$--0.35} (see Table~\ref{tabcomp}), and find that $\left<\hbox{\rm Pab}\right>$ is the
same within the standard deviations of the distributions.  Again,
the mean properties of the different RRab samples seem virtually
indistinguishable from this statistical point of view. \par

\begin{figure}[!ht]
\begin{center}
\includegraphics[width=1.0\textwidth]{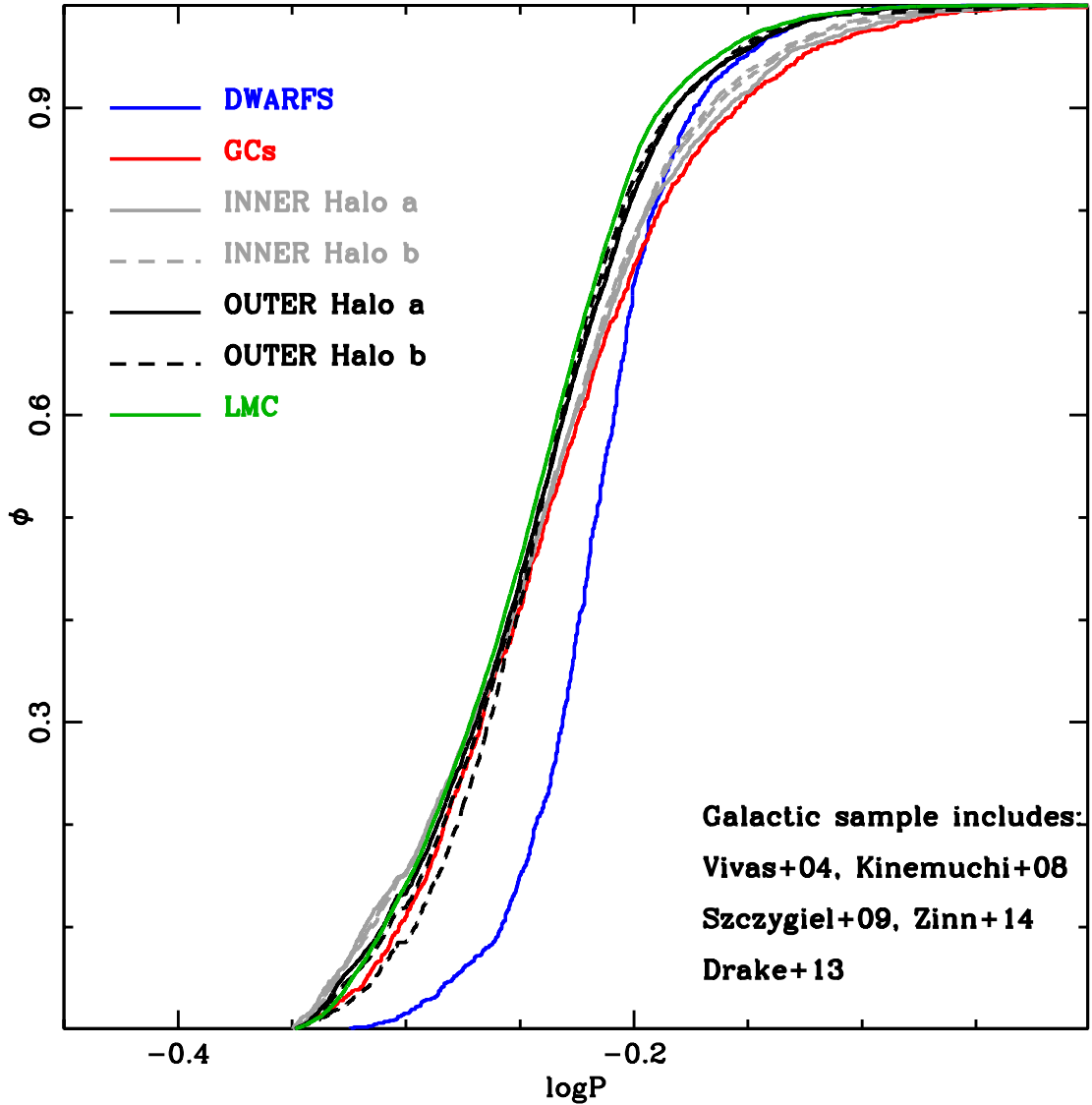}
\caption{Cumulative period distributions for RRLs used to compute the
  likelihood that different samples are drawn from the same parent
  population. The data collected in
  this study represent: inner halo a (gray solid ine), inner halo b
  (gray dashed line), outer halo a (black solid line), outer halo b
  (black dashed line), classical dSph plus ultra-faint dwarf galaxies
  (blue solid line), GCs (red solid line) and LMC (green solid line).  
\label{fig12}  
}
\end{center}
\end{figure}

To compare these distributions in more rigorous detail, we have
performed two more sophisticated statistical analyses.  The first
is a chi-squared test on the histograms smoothed with a Gaussian
filter (see blue solid curves in Fig.~\ref{fig11}), and the second
is a Kolmogorov-Smirnov test applied to the cumulative period
distributions (see Fig.~\ref{fig12}).  For the purposes of these
last two tests, only, we take advantage of the large sample of RRLs in
the Large Magellanic Cloud that has been provided by the OGLE
project \citep[17,693 stars;][]{soszynski09a}.  Although the RRLs in the LMC have been surveyed in
the $I$ band by OGLE, we decided to attempt to match the amplitude
bias of the CATALINA sample by using only those stars with A$_I
\ge$ 0.26 mag. This corresponds to a scale factor $A_V/A_I = 1.58$ as
found by \cite{dicriscienzo11}.  These two statistical
tests returned very similar results, which are listed in
Table~\ref{tabprob} and can be summarized as follows: 

\begin{itemize}
\item Inner halo a and b have a likelihood of 30\% of being drawn from
  the same parent populations. This likelihood increases dramatically
  to 90\% when only the CATALINA survey is considered. As shown in
  Fig.~\ref{fig11}, these two regions show broad period distributions
  including the longest-period RRLs.

\item Outer halo a and b do not share a common period
  distribution, with each other or with the inner halo. Each component of the
  outer halo seems to have a more peaked distribution than that of the inner halo.
  In particular, outer halo b seems to show a relative deficiency
  of RRab stars with high amplitudes and short periods (hereinafter, ``HASP''). This can not be
  attributed to a completeness effect because large amplitudes are 
  the easiest ones to detect.

\item The GC period distribution mildly resembles that of the RRLs
  in the inner halo (within $\sim\,$14 kpc, a$+$b). This is reasonable
  when we consider that 23 out of the 35 GCs considered here belong
  to this region. We note that this result must be viewed as
  preliminary because: 1) the GC period distribution strongly
  depends on the choice of GCs included in the full catalog; 2) GCs
  have their own space motions that may eventually tell us more
  about the halo region to which they belong, rather than where they
  happen to be currently located.

\item The UFD$+$dSph distribution is quite different from the
others. It is very peaked around the mean period (see
Table~\ref{tabcomp}). In particular, this distribution is quite
lacking in HASP RRab stars compared to the others: {\it
none\/}\footnote{For completeness, we should mention that there
are two RR Lyrae variables in the HASP region of the Bailey
diagram for the dSph galaxy Cetus: V11 and V173 in
\cite{bernard09}.  However, both these stars are peculiar: they are
subluminous and have much shorter periods
than any other RRab stars in this galaxy, and moreover the light curve for V173 is of very
poor quality.  For these reasons we have not included these
two stars in the UFD+dSph histogram in Fig.~\ref{fig11}, but they are
included in the Cetus panel of Fig.~\ref{fig9}.} of the RRLs in dwarfs with high
amplitude ($A_V\ge 1.0$) reaches periods P$\,\lsim\,0.48\,$d
(log~P$\,\lsim\,$--0.32).  Fundamental-mode RR Lyraes with large
amplitudes and periods less than 0.48$\,$d are quite common both
in globular clusters and in the halo field.  Even the {\it
relative\/} deficiency of HASP RRab stars in outer halo b already
noted above is not as complete as in the UFD$+$dSph sample.  It is
true that at the other extreme, UFDs are seen to help in filling
out the long-period tail (\cf\ Fig.~\ref{fig9}a) but---within the
limited statistics available---they do not appear to contribute to
the short-period tail.  

\item We find a very low formal likelihood that the LMC sample
  matches any of the others (see Table~\ref{tabprob}), as is also
  true for classical dSphs and UFDs. We note here, though, that this
  result may be affected by the very high temporal and photometric
  completeness of the OGLE sample for the LMC. In other words,
  when huge samples are available (as they are here for the LMC and the Milky Way halo), any minor
  difference in selection biases achieves very high statistical significance and can be
  mistaken for a real physical difference between the samples.  A
  fair comparison with the LMC sample needs much more complete RRL surveys
  of the Galactic halo. The same consideration is relevant to the recent result
  discussed by \citet{zinn14} where, on the basis of the OGLE
  \citep{soszynski09a} and their own QUEST RRL samples, the
  authors declined to exclude the possibility that the smooth Galactic halo may have formed
  from a combination of LMC- and Sagittarius-like galaxies.  We
  believe---but are unable to prove---that because our comparisons
  between the halo, the GCs, and UFD+dSphs are based on samples closer
  in size, and probably with selection effects more nearly similar
  to each other than to the OGLE LMC sample, our conclusions regarding
  the statistical significance of their differences are probably more
  reliable than similar conclusions relating to the LMC.  
\end{itemize}

%%%%%%%%%%%%%%%%%%%%%%%%%%%%%%%%%%%%%%%%%%%%%%%%%%%%%%%%%%%%%%%%%%%%
\section{Discussion: Building up the Galactic Halo with dwarf galaxies}\label{concl}
%%%%%%%%%%%%%%%%%%%%%%%%%%%%%%%%%%%%%%%%%%%%%%%%%%%%%%%%%%%%%%%%%%%%

Perhaps it is worthwhile to remind the reader that the dwarf
galaxies in the Milky Way halo contain stellar populations with
some range of ages and metallicities.  The classical globular
clusters as a class exhibit a {\it broad\/} range of metallicities
($-2.5 < \hbox{\rm [Fe/H]} < 0.0$, roughly speaking) and a rather
limited range of age.  The Milky Way field halo also clearly has a
broad range of metal abundances, but we have less information as
to its range of age.  The range of metal abundance in the dwarf
galaxies is somewhat narrower than among the GCs,
[Fe/H]$\,\lsim\,$--1 or so, but some---including \leoi, Carina,
and Fornax, for example---have very broad ranges of age, much
greater than the range found among the GCs.  

However, it must also be remembered that the range of age and
metal abundance that is conducive to the formation of RR Lyraes is
more restricted: the most metal-rich globular clusters contain no
RR Lyraes, and the most metal-poor contain few, while moderately
metal-poor clusters contain the most.  Populations younger than
$\sim 10\,$Gyr probably cannot produce RR Lyraes by the normal
evolution of single stars.  The range of metallicities where the
globular clusters and the field halo are proficient at producing
RR Lyraes, $-1.0 \lsim \hbox{\rm [Fe/H]} \lsim -2.0$, is very well
represented in the dwarf galaxies, and, as we have just mentioned,
stars in the dwarf galaxies have---if anything---a broader range
of age than those in the globular clusters and the field halo. 
That being the case, it is remarkable that there exists a class of
RR Lyrae star that is {\it present\/} in the globular clusters and
the field halo and {\it absent\/} in the dwarf galaxies.  One would
think that if there were some odd corner of age-metallicity space
that is capable of making a particular class of RR Lyrae stars,
but is not populated among the globular clusters, that niche would
be more likely to be occupied in the dwarf galaxies: if anything,
the dwarf galaxies should have {\it more\/} varieties of RR Lyrae
stars than the globular clusters, not {\it fewer\/}.  

One possible exception to this rule might be if the HASP RR Lyrae
stars were uniquely the progeny of stellar collisions or merged
gravitational-capture binaries; such stars could then be present
in the globular clusters and not in the much sparser dwarf
galaxies.  But if this is the case, then why are these ``special''
RR Lyraes also found in the field halo, in roughly the same
relative numbers as in the globular clusters?  To be consistent,
this explanation would suggest that a major fraction of the field
halo stars came from globular clusters that were disrupted {\it
recently\/}, i.e., only {\it after\/} the globular clusters had
had enough time to produce the usual numbers of stellar collisions
and gravitational-capture binaries.  But even this seems highly
unlikely, since halo field red giants do not show the light-element
(C, N, O, Na, Mg, Al, etc.) correlations and anti-correlations found
among the globular cluster giants.

In conclusion, and for the sake of argument, let us assume that in
the early days of the Galactic halo, star formation occurred in one
fundamentally indistinguishable family of structures, ranging in
mass, that all underwent a common process of internal chemical
evolution.  During this time the young main sequence stars that
grew into today's RR Lyraes were born.  During subsequent
dynamical evolution most of these structures dissolved or were
disrupted, releasing the globular clusters and field stars of
today's Galactic halo.  At the same time a few of those
structures---by random chance---survived, continued evolving
internally, and became the classical dSph and UFD galaxies that we
see today.  

It seems that on the basis of the available RRL samples we can say
that this scenario is not supported. Neither the inner
(d$_G\,\lsim\,$14 kpc) nor the outer (d$_G\,\gsim\,$14 kpc)
regions of the Galactic halo, nor any linear combination of the
two, can be formed simply from the progenitors of today's
classical dSph or UFD galaxies. If present in sufficient numbers,
structures like those that became today's UFDs may have
contributed to the long-period tail of the Galactic halo RRL period
distribution, but---barring a major statistical
fluke---neither they nor the dSph progenitors appear to have been
capable of producing the kind of main sequence stars that became the
HASP RRab variables now found in significant numbers throughout the
Galactic halo and in the globular clusters.  

Unless it can be shown that a protracted residence within the
environment of a dSph or UFD galaxy can somehow alter the internal
structure and evolution of an isolated star between the main
sequence and RR Lyrae phases of its life, it seems that the simple
model---where the dSphs and UFDs are surviving examples of {\it
representative\/} proto-halo fragments---is disfavored.  Instead
it seems that already at the time when the future RR Lyrae stars
were being born, the milieux that were halo-to-be or
UFD/dSph-to-be were already different in some way.  In other
words, the progenitors of the Milky Way halo and of the surviving dwarf
galaxies appear to have been different, non-representative
subsamples of the proto-Galactic material, perhaps coming
preferentially from distinct ranges of mass or kinematic
properties, possessing different typical ratios of baryonic to
dark matter, or being distinguished in some other relevant
characteristic.  Furthermore, the RRLs within the Galactic halo
itself seem to have different period distributions when our consideration moves
from the inner to the outer regions. This argues that the halo is
itself formed from at least two different classes of progenitor or
distinct, non-representative subsamples of a continuum of
progenitors \citep[see also][]{chiappini01}, as confirmed by
kinematical surveys \citep{carollo07,kinman12}.  

As was pointed out by an astute referee, there is now good
evidence that all the known dSph and UFD galaxies---except
Hercules---and the ``young halo'' globular clusters are
distributed on a thin, rotationally supported planar structure
containing the Galactic poles \citep[the VPOS: Vast POlar
Structure;][] {pawlowski12,pawlowski13}.  It seems far more likely
that the VPOS and the distinct star systems located within it
result from the disruption of a single infalling object, rather
than from the isotropic accretion of a large number of smaller
structures.  This interpretation is inconsistent with the
predictions of classical $\Lambda$CDM models for the formation of
the halo in its entirety, but is consistent with our inference
that the progenitor(s) of the UDF and dSph galaxies were different
from the progenitor(s) of the Milky Way field halo stars and
normal globular clusters. 

As a final remark, since GCs represent only a small fraction of
the Galactic halo, and because we now recognize that their
histories were probably more complicated than we used to
think---and are still not really understood---we would like to
suggest that their continued use as representative tracers of the
Galactic halo is not necessarily preferable to other comparisons
that are now becoming possible.

We plan to extend this work to other dwarf spheroidal and
irregular galaxies that are being studied within the homogeneous
photometry project; we expect this will provide an opportunity to
further test our conclusions.

%%%%%%%%%%%%%%%%%%%%%%%%%%%%%%%%%%%%%%%%%%%%%%%%%%%%%%%%

\acknowledgments
The authors sincerely thank the Instituto de Astrof\'isica de
Canarias, where the main conclusions of this paper were drawn and
our collaboration was strengthened. Financial support for this
work was provided by the IAC (grant 310394) and the Education and
Science Ministry of Spain (grant AYA2010-16717).  PBS is pleased
to acknowledge financial support from the Severo Ochoa Program,
sponsored by the Spanish Ministry of Science and Education, and
from the Erasmus Mundus-AstroMundus Consortium of Universities. GF
has been partially supported by the Futuro in Ricerca 2013 (RBFR13J716) and by the project
Cosmic-Lab (http://www.cosmic-lab.eu) funded by the European
Research Council under contract ERC-2010-AdG-267675.  This work
was partially supported by PRIN--INAF 2011 ``Tracing the formation
and evolution of the Galactic halo with VST'' (P.I.: M.~Marconi)
and by PRIN--MIUR (2010LY5N2T) ``Chemical and dynamical evolution
of the Milky Way and Local Group galaxies'' (P.I.: F.~Matteucci). 
Howard Bond, Ed Olszewski, and Nick Suntzeff have contributed
proprietary data to the homogeneous photometry project, some of
which was used here.

%%%%%%%%%%%%%%%%%%%%%%%%%%%%%%%%%%%%%%%%%%%%%%%%%%%%%%%%

%%%%%%%%%%%%%%%%%%%%%%%%%%%%%%%%%%%%%%%%%%%%%%%%%%%%%%%%%%%%%%%%%%%%
\section{APPENDIX: Previously identified stars}\label{appen}
%%%%%%%%%%%%%%%%%%%%%%%%%%%%%%%%%%%%%%%%%%%%%%%%%%%%%%%%%%%%%%%%%%%%

Leo~I was previously surveyed for bright variable stars by
\cite{hodgewright78}, who identified candidate Anomalous Cepheids
and RR Lyraes, and by \cite{menzies02}, who identified candidate
long-period variables.  

Hodge \& Wright did not publish coordinates for their stars, but
they did provide a finding chart.  We extracted a JPEG version of
their finding chart from the on-line electronic edition of their
article and converted it to FITS format.  Using a slightly
modernized version of the \cite{stetson79} software for astrometry
and photometry from digitized photographic plates, we measured the
positions of stars within the Hodge \& Wright finding chart.  We
then used quadratic equations to transform those positions to the
coordinate system of our Leo~I catalog.  We were able to match 355
stars within a tolerance of 4\sec\ (the r.m.s.\ residuals were
0\Sec45 in right ascension and 0\Sec37 in declination).  We were
then able to geometrically transform their finding chart and overlay it on our
stacked digital image of Leo~I.  Overall, we feel we are able to
recover positions from the Hodge \& Wright finding chart with a
precision better than 0\Sec5.  We considered stars within several
arcseconds of each of the positions indicated in the published
chart, basing our identification on (in order of decreasing
importance) (1)~evidence of variability in our data, (2)~relative
proximity to the indicated position, and (3)~similarity of the mean
$B$-band magnitude.  We summarize our resulting
cross-identifications in Tables~\ref{tabrr} (RR Lyrae stars),
\ref{tabac} (AC/spC stars) and \ref{hodgewright} (stars whose variability
we are unable to confirm).

Notes on individual stars:
\begin{itemize}
\item The tick marks for HW1 in their chart enclose nothing but blank sky in our
image.  However, there are two of our variable candidates within 5\sec\ of this
position:  V66 is a candidate long-period variable
with $\left<B\right> \approx 22.1$ lying 4\Sec2 from the predicted
position, and V59 is an AC/spC with P$\,=\,1.49328\,$d,
$\left<B\right> \approx 20.7$ lying 4\Sec8 from the predicted
position.  Despite its greater offset, the latter star more
closely resembles HW1 and we accept it as the match.  

\item HW2 has no detectable star in our image within the tick marks on
the finding chart.  There is a spot of fuzz most likely
representing a galaxy about 1\sec\ southwest of the indicated
position, but this is probably too faint to have been detected in
Hodge \& Wright's photographic material.  One of our AC/spC
candidates does lie almost 6\sec\ ($> 10\sigma$) from the indicated position. 
Since it has approximately suitable period and magnitude,
we provisionally make this identification despite the large offset.  

\item HW20 has at least eleven stars lying closer to the indicated position than
our adopted match, but the one we have chosen is clearly the most suitable
on the basis of its mean $B$-band magnitude, and furthermore its
Welch/Stetson variability index is not far below the
detection threshold that we had imposed.  This star may be variable, but we
are not confident enough to claim definite confirmation.
\end{itemize}

\cite{menzies02} did not publish a finding chart for their five
stars in Leo~I.  They did publish a table of equatorial
coordinates for five luminous red stars in the galaxy field, but
unfortunately we have not been able to find any relationship
whatsoever between those coordinates and reality: we have marked
their coordinates on our digital image of Leo~I and we have also
marked all the stars that are both luminous and red in our
catalog, and we are not able to recognize any correspondence
between the two patterns.  For example, their stars A and B
constitute a pair separated by 29\Sec7, slightly inclined with
respect to the east-west direction.  We have examined our list of
luminous red stars for pairs separated by $\sim$30\sec, slightly
inclined to the east-west direction, and do find a few.  But when we
accept any of those provisional matches none of the other three stars
from Menzies et al.\ coincides with anything interesting.  

\cite{alw86} published a finding chart identifing several
luminous, red stars in Leo~I, and we give cross-identifications
with our catalog in Table~\ref{alw}.

%-------------------------------------------------------------------
\bibliographystyle{aa} 
%\bibliography{ms} 

%%%%%%%%%%%%%%%%%%%%%%%%%%%%%%%%%%%%%%%%%%%%%%%%%%%%%%%%%%%%%%%%%%%%%%%%%%
%%%%%%%%%%%%%%%%%%%%%%%%%%%%%%%%%%%%%%%%%%%%%%%%%%%%%%%%%%%%%%%%%%%%%%%%%%
%%%%%%%%%%%%%%%%%%%%%%%%%%%%%%%%%%%%%%%%%%%%%%%%%%%%%%%%%%%%%%%%%%%%%%%%%%
%%%%%%%%%%%%%%%%%%%%%%%%%%%%%%%%%%%%%%%%%%%%%%%%%%%%%%%%%%%%%%%%%%%%%%%%%%

\clearpage
%%%%%%%%%%%%%%%%%%%%%%%%%%%%%%%%%%%%%%%%%%%%%%%%%%%%%%%%%%%%%%%%%%%%%%%%%%
%TAB 1
%%%%%%%%%%%%%%%%%%%%%%%%%%%%%%%%%%%%%%%%%%%%%%%%%%%%%%%%%%%%%%%%%%%%%%%%%%
\begin{deluxetable}{llllcccccl} 
\tabletypesize{\tiny}
\tablewidth{0pt}
\tablenum{1}
%\centering
\tablecaption{\leoi\/ Observations}
\label{obs}
\tablehead{
\colhead{Run ID}    &
 \colhead{Dates} &
 \colhead{Telescope} &
 \colhead{Camera/Detector} &
 \colhead{$U$} &
 \colhead{$B$} &
 \colhead{$V$} &
 \colhead{$R$} &
 \colhead{$I$} &
 \colhead{}
}
\startdata
\ 1 nbs      & 1983 01 08-11    & CTIO 4.0m        & RCA                      & -- & 13 & 17 &  4 & -- & \\
\ 2 p200     & 1986 01 19-20    & Palomar 5.1m     & 4-Shooter                & -- & -- & 17 & 11 & -- & \\
\ 3 f8       & 1986 02 19       & CTIO 4.0m        & RCA                      & -- & 12 & -- & -- & -- & \\
\ 4 km       & 1986 05 10-14    & INT 2.5m         & RCA                      & -- &  5 &  2 &  4 &  1 & \\
\ 5 f10      & 1989 03 07       & CTIO 4.0m        & TI1                      & -- &  6 &  6 &  3 & -- & \\
\ 6 f9       & 1989 03 15       & CTIO 0.9m        & RCA5                     & -- &  4 &  4 &  4 & -- & \\
\ 7 f16      & 1992 03 06       & CTIO 4.0m        & Tek1K-2                  & -- &  5 &  5 & -- & -- & \\
\ 8 demers   & 1992 03 26-27    & CFHT 3.6m        & Lick2                    & -- &  7 &  7 & -- & -- & \\
\ 9 schommer & 1992 05 31       & CTIO 0.9m        & Tek1K-1                  & -- & -- &  2 & -- & -- & \\
10 f15b      & 1994 02 06       & CTIO 0.9m        & Tek1024                  & -- & -- &  1 & -- &  1 & \\
11 f15c      & 1994 03 13       & CTIO 0.9m        & Tek1024                  & -- & -- &  1 & -- &  1 & \\
12 bond      & 1994 12 03       & KPNO 4.0m        & t2kb                     &  3 &  3 &  3 & -- &  3 & \\
13 feb95     & 1995 02 04-05    & Steward Bok 2.3m & 12x8bok                  & -- & -- & 72 & -- & -- & \\
14 mdm(a)    & 1995 02 27       & Hiltner 2.4m     & Charlotte Tek 1024$^2$   & -- & -- &  4 & -- & -- & \\
15 jka       & 1995 03 24       & INT 2.5m         & TEK3                     & -- & -- &  1 & -- & -- & \\
16 apr95     & 1995 04 29-05 01 & Steward Bok 2.3m & 12x8bok                  & -- & -- & 26 & -- & -- & \\
17 mdm(b)    & 1995 05 05       & Hiltner 2.4m     & Charlotte Tek 1024$^2$   & -- & -- &  4 & -- & -- & \\
18 bond2     & 1996 03 13       & KPNO 4.0m        & t2kb                     &  1 &  1 &  1 & -- &  1 & \\
19 int       & 1998 06 24       & INT 2.5m         & WFC EEV42                & -- &  1 &  1 & -- &  1 & $\times\,$4 \\
20 wfi3      & 1999 03 17       & ESO 2.2m         & WFI                      & -- & -- & 13 & -- & 13 & $\times\,$8 \\
21 emmi      & 1999 04 14-15    & ESO NTT 3.6m     & EMMI                     & -- &  6 &  6 & -- &  1 & \\
22 fors9912  & 1999 12 02       & ESO VLT 8.0m     & FORS1                    & -- & -- &  3 & -- &  3 & \\
23 wfi4      & 2000 04 21-26    & ESO/MP 2.2m      & WFI                      & -- & 13 & 31 & -- &  2 & $\times\,$8 \\
24 wfi       & 2000 04 22       & ESO/MP 2.2m      & WFI                      & -- &  4 & 10 &  6 &  1 & $\times\,$8 \\
25 tng       & 2001 01 24-26    & TNG 3.6m         & LRS Loral                & -- & 30 & 49 & -- & -- & \\
26 bellaz    & 2001 03 19-22    & TNG 3.6m         & LRS Loral                & -- & -- &  8 & -- &  8 & \\
27 suba      & 2001 03 20       & Subaru 8.2m      & SuprimeCam               & -- & -- & -- &  7 & -- & $\times\,$8 \\
28 suba2     & 2001 03 20-21    & Subaru 8.2m      & SuprimeCam               & -- & -- & 28 & 16 & -- & $\times\,$8 \\
29 fors0204  & 2002 04 13       & ESO VLT 8.0m     & FORS2 MIT/LL mosaic      & -- & -- &  4 & -- &  4 & $\times\,$2 \\
30 abi36     & 2002 11 13-15    & KPNO 0.9m        & s2kb                     & -- &  3 &  3 &  3 &  3 & \\
31 lbt(a)    & 2006 11 21-25    & LBT 8.4m         & LBC                      & 43 & 20 & -- & -- & -- & $\times\,$4 \\
32 lbt(b)    & 2006 12 16       & LBT 8.4m         & LBC                      & -- & -- &  5 & -- & -- & $\times\,$4
\enddata
\tablecomments{
 1. Observer N.~Suntzeff;
 2. Observer R.~Schommer?;
 3. Observer N.~Suntzeff or R.~Schommer?;
 4. Observer ``KM'';
 5. Observer N.~Suntzeff or R.~Schommer?;
 6. Observer N.~Suntzeff or R.~Schommer?;
 7. Observer N.~Suntzeff or R.~Schommer?;
 8. Observer D.~Crampton and J.~Hutchings;
 9. Observer R.~Schommer;
10. Observer N.~Suntzeff or R.~Schommer?;
11. Observer N.~Suntzeff or R.~Schommer?;
12. Observer H.~E.~Bond;
13. Observer N.~Suntzeff or R.~Schommer?;
14. Observer unknown;
15. Observer ``JKA'';
16. Observer N.~Suntzeff or R.~Schommer?;
17. Observer unknown;
18. Observer H.~E.~Bond;
19. Observer unknown;
20. Program ID unknown, observer unknown;
21. Program ID 63.H-0365(A), observer unknown;
22. Program ID 64.N-0421(A), observer unknown;
23. Program ID unknown, observer unknown;
24. Program ID 065.O-0530, PI Saviane, observer Momany;
25. Observer Bono;
26. Observer ``bellaz'';
27. Proposal ID o00103, observers Arimoto, Ikuta, and Jablonka;
28. Proposal ID o00103, observers Arimoto, Ikuta, and Jablonka;
29. Program ID 69.D-0455(B), observer unknown;
30. Proposal ID ``Saha'', observers Dolphin and Thim;
31. Observer unknown;
32. Observer unknown.
}
\end{deluxetable}
\clearpage

%%%%%%%%%%%%%%%%%%%%%%%%%%%%%%%%%%%%%%%%%%%%%%%%%%%%%%%%%%%%%%%%%%%%%%%%%%
% TAB2
%%%%%%%%%%%%%%%%%%%%%%%%%%%%%%%%%%%%%%%%%%%%%%%%%%%%%%%%%%%%%%%%%%%%%%%%%%
\begin{deluxetable}{lll cccc cccc c cc}
\centering
\tabletypesize{\tiny}
\tablewidth{0pt}
\tablenum{2}
\tablecaption{RR~Lyrae candidates: periods, photometric parameters and
  positions. }
\label{tabrr}
\tablehead{
\colhead{ID} &
\colhead{Var} &
\colhead{period} &
\colhead{ $\left<B\right>$} &
\colhead{ $\left<V\right>$} &
\colhead{ $\left<R\right>$} &
\colhead{ $\left<I\right>$} & 
\colhead{$A_B$}  & 
\colhead{$A_V$}  & 
\colhead{$A_R$}  & 
\colhead{$A_I$}  & 
\colhead{Q} &
\colhead{RA} &
\colhead{Dec} \\
	\colhead{}&
	\colhead{type}&
	\colhead{d}&
	\colhead{mag}&
	\colhead{mag}&
	\colhead{mag}&
	\colhead{mag}&
	\colhead{mag}&
	\colhead{mag}&
	\colhead{mag}&
	\colhead{mag}&
	\colhead{}&
	\colhead{hh mm ss}&
	\colhead{dd mm ss}
\\
}
\startdata
V1*       & RRab    & 0.515750   & 23.72 & 22.79 & 21.60 &  ---  &  --- & 1.62 &  --- &  --- & C & 10 07 13.92 & +12 21 50.8 \\
V2*       & RRab    & 0.627407   & 22.99 & 22.53 & 22.14 &  ---  & 1.14 & 1.17 & 0.96 &  --- & A & 10 07 18.89 & +12 13 10.2 \\
V3*       & RRab    & 0.554130   & 23.04 & 22.61 & 22.31 &  ---  & 0.84 & 1.03 &  --- &  --- & B & 10 07 28.27 & +12 14 40.4 \\
V4*       & RRab    & 0.638565   & 22.88 & 22.53 & 22.50 &  ---  & 1.01 & 0.93 &  --- &  --- & B & 10 07 33.35 & +12 21 56.1 \\
V5*       & RRab    & 0.593564   & 23.25 & 22.75 & 22.46 & 22.16 & 1.10 & 0.95 & 0.60 &  --- & B & 10 07 36.68 & +12 18 49.1 \\
V6*       & RRab    & 0.698670   & 23.13 & 22.61 & 22.28 & 21.74 & 1.00 & 0.88 & 0.75 &  --- & C & 10 07 38.33 & +12 12 40.3 \\
V7*       & RRab    & 0.6024018  & 23.04 & 22.75 & 22.57 & 22.25 & 0.96 & 0.79 & 0.23 &  --- & A & 10 07 40.62 & +12 18 55.6 \\
V8*       & RRab    & 0.537169   & 23.15 & 22.59 & 22.20 & 22.15 & 1.14 & 0.94 &  --- &  --- & B & 10 07 44.43 & +12 13 07.5 \\
V9        & RRab    & 0.574817   & 22.81 & 22.69 & 22.65 & 22.34 & 1.17 & 1.22 & 0.36 &  --- & A & 10 07 44.44 & +12 17 17.4 \\
V10       & RRab    & 0.681396   & 23.00 & 22.61 & 22.41 &  ---  & 0.45 & 0.37 & 0.32 &  --- & B & 10 07 46.57 & +12 20 15.0 \\
V11       & RRab    & 0.515261   & 23.05 & 22.64 & 22.44 & 22.45 & 1.24 & 1.18 &  --- &  --- & B & 10 07 48.43 & +12 18 56.8 \\
V13       & RRab    & 0.636654   & 23.13 & 22.64 & 22.47 &  ---  & 0.58 & 0.46 & 0.40 &  --- & A & 10 07 52.04 & +12 15 34.6 \\
V14       & RRab    & 0.708036   & 23.15 & 22.90 & 22.70 & 22.44 & 0.84 & 0.60 & 0.18 &  --- & C & 10 07 55.61 & +12 20 53.9 \\
V15*      & RRab    & 0.630248   & 23.10 & 22.72 & 22.56 &  ---  & 0.79 & 0.68 & 0.38 &  --- & A & 10 07 55.99 & +12 14 13.6 \\
V16       & RRab    & 0.601348   & 23.04 & 22.78 & 22.35 &  ---  & 1.00 & 0.90 & 0.68 &  --- & A & 10 07 56.41 & +12 19 47.1 \\
V17       & RRc     & 0.3131235  & 23.00 & 22.74 & 22.36 &  ---  & 0.72 & 0.49 &  --- &  --- & A & 10 07 56.48 & +12 13 54.4 \\
V18       & RRab    & 0.628029   & 23.02 & 22.70 & 22.41 &  ---  & 1.05 & 0.79 & 0.45 &  --- & A & 10 07 56.55 & +12 18 16.1 \\
V19       & RRab    & 0.607947   & 23.08 & 22.66 & 22.39 &  ---  & 0.90 & 0.83 & 0.34 &  --- & A & 10 07 58.54 & +12 18 00.0 \\
V21       & RRab    & 0.659662   & 22.88 & 22.67 & 22.56 & 22.04 & 0.98 & 0.85 & 0.37 &  --- & B & 10 08 00.69 & +12 24 54.3 \\
V22       & RRab    & 0.557603   & 23.00 & 22.70 & 22.50 & 22.19 & 1.03 & 0.97 & 0.42 &  --- & B & 10 08 01.08 & +12 17 54.0 \\
V23       & RRc     & 0.2737468  & 23.05 & 22.69 & 22.54 & 22.16 & 0.45 & 0.37 & 0.32 &  --- & B & 10 08 01.44 & +12 20 09.2 \\
V24*      & RRab    & 0.607729   & 23.20 & 22.78 & 22.57 & 22.25 & 0.66 & 0.76 & 0.30 &  --- & A & 10 08 02.45 & +12 18 07.4 \\
V25       & RRab    & 0.5615871  & 23.11 & 22.68 & 22.54 &  ---  & 0.84 & 1.03 &  --- &  --- & A & 10 08 03.48 & +12 19 58.0 \\
V26       & RRab    & 0.541359   & 23.02 & 22.64 &  ---  & 9.999 & 1.16 & 1.13 &  --- &  --- & B & 10 08 04.08 & +12 15 57.6 \\
V28       & RRc     & 0.385380   & 22.94 & 22.67 & 22.49 & 22.12 & 0.72 & 0.44 & 0.38 &  --- & B & 10 08 05.74 & +12 19 24.9 \\
V31       & RRab    & 0.6026533  & 23.07 & 22.77 & 22.49 &  ---  & 1.19 & 0.96 & 0.70 &  --- & A & 10 08 07.07 & +12 21 32.5 \\
V32       & RRab    & 0.580355   & 22.97 & 22.66 & 22.44 & 22.27 & 1.05 & 0.88 & 0.51 &  --- & A & 10 08 07.68 & +12 18 50.1 \\
V35       & RRab    & 0.626002   & 23.08 & 22.73 & 22.57 &  ---  & 0.57 & 0.64 & 0.22 &  --- & A & 10 08 08.25 & +12 13 48.6 \\
V36       & RRab    & 0.558055   & 23.07 & 22.70 & 22.75 & 22.23 & 0.72 & 0.85 & 0.14 &  --- & A & 10 08 08.27 & +12 15 13.0 \\
V37*      & RRab    & 0.5865854  & 22.96 & 22.74 & 22.66 & 22.10 & 1.04 & 1.08 & 0.78 &  --- & B & 10 08 08.52 & +12 13 15.9 \\
V38       & RRab    & 0.5694029  & 22.94 & 22.66 & 22.40 & 21.92 & 1.05 & 0.90 & 0.38 &  --- & A & 10 08 08.82 & +12 21 23.8 \\
V42       & RRab    & 0.590415   & 22.89 & 22.59 & 22.44 & 21.91 & 0.93 & 0.84 & 0.26 &  --- & A & 10 08 10.01 & +12 16 16.7 \\
V43       & RRab    & 0.518913   & 23.11 & 22.65 & 22.01 & 21.98 & 0.70 & 0.60 & 0.50 &  --- & B & 10 08 10.30 & +12 15 14.9 \\
V44       & RRab    & 0.5368825  & 22.82 & 22.53 & 22.50 & 21.84 & 1.32 & 0.96 &  --- &  --- & A & 10 08 10.75 & +12 18 28.9 \\
V47       & RRc     & 0.3741098  & 22.89 & 22.59 & 22.42 & 22.22 & 0.43 & 0.40 & 0.31 &  --- & A & 10 08 11.42 & +12 17 55.8 \\
V49       & RRab    & 0.6106710  & 23.11 & 22.79 & 22.48 & 22.05 & 0.85 & 0.63 & 0.48 &  --- & A & 10 08 12.23 & +12 20 50.1 \\
V51       & RRab    & 0.631211   & 23.09 & 22.62 & 22.46 & 21.98 & 0.38 & 0.43 & 0.27 &  --- & B & 10 08 12.36 & +12 18 53.0 \\
V52       & RRab    & 0.5576776  & 22.99 & 22.66 & 22.57 & 22.14 & 1.30 & 1.03 & 0.40 &  --- & A & 10 08 12.45 & +12 17 09.2 \\
V53       & RRab    & 0.5541654  & 22.87 & 22.54 & 22.09 & 22.01 & 1.23 & 0.94 & 0.54 &  --- & A & 10 08 12.53 & +12 17 54.7 \\
V56       & RRc     & 0.3591109  & 22.82 & 22.56 & 22.44 & 22.08 & 0.58 & 0.48 & 0.44 &  --- & A & 10 08 13.29 & +12 16 41.7 \\
V57=HW14  & RRab    & 0.5644922  & 22.91 & 22.42 & 22.06 & 21.74 & 0.78 & 0.54 & 0.44 &  --- & A & 10 08 14.05 & +12 15 06.2 \\
V58*      & RRc     & 0.368497   & 22.86 & 22.58 & 22.59 & 22.15 & 0.71 & 0.58 & 0.24 &  --- & A & 10 08 14.18 & +12 15 30.2 \\
V59       & RRab    & 0.635755   & 22.87 & 22.41 & 22.04 & 21.71 & 0.62 & 0.40 & 0.32 & 0.25 & A & 10 08 14.20 & +12 17 38.2 \\
V60       & RRc     & 0.3755769  & 22.95 & 22.69 & 22.54 & 22.24 & 0.55 & 0.43 & 0.38 &  --- & A & 10 08 14.30 & +12 15 17.5 \\
V63       & RRab    & 0.517857   & 22.90 & 22.73 & 22.77 & 22.09 & 1.52 & 1.20 &  --- &  --- & B & 10 08 15.37 & +12 17 32.4 \\
V64       & RRab    & 0.5419958  & 23.00 & 22.77 & 22.41 & 22.19 & 1.14 & 1.06 & 0.67 &  --- & A & 10 08 15.38 & +12 15 09.6 \\
V65       & RRab    & 0.5252377  & 22.89 & 22.55 & 22.49 & 22.36 & 1.38 & 1.09 & 0.40 &  --- & A & 10 08 15.52 & +12 16 27.9 \\
V66       & RRab    & 0.5413377  & 23.05 & 22.79 & 22.30 & 21.94 & 1.32 & 1.35 & 1.28 &  --- & A & 10 08 15.70 & +12 16 51.8 \\
V67       & RRab    & 0.5505050  & 23.08 & 22.76 & 22.44 & 22.11 & 1.19 & 0.95 & 0.57 &  --- & A & 10 08 16.12 & +12 17 52.5 \\
V68       & RRab    & 0.5348781  & 22.82 & 22.63 & 22.67 & 22.23 & 1.76 & 1.22 & 0.90 &  --- & A & 10 08 16.32 & +12 18 24.3 \\
V70       & RRab    & 0.5559762  & 23.02 & 22.78 & 22.41 & 22.43 & 1.16 & 0.73 & 0.53 &  --- & A & 10 08 16.44 & +12 18 12.1 \\
V72       & RRab    & 0.601951   & 23.10 & 22.78 & 22.47 & 22.27 & 0.57 & 0.54 & 0.29 &  --- & A & 10 08 16.79 & +12 22 03.7 \\
V73       & RRab    & 0.547411   & 22.95 & 22.63 & 22.75 &  ---  & 1.46 & 1.18 & 0.80 &  --- & A & 10 08 17.01 & +12 13 37.0 \\
V74       & RRab    & 0.5097438  & 22.98 & 22.71 & 22.35 & 22.33 & 2.17 & 1.27 & 0.54 &  --- & A & 10 08 17.44 & +12 22 04.6 \\
V76       & RRab    & 0.548736   & 23.01 & 22.71 & 22.45 & 22.37 & 1.16 & 1.11 & 0.42 &  --- & A & 10 08 17.59 & +12 15 25.1 \\
V77       & RRab    & 0.5883625  & 22.99 & 22.67 & 22.62 & 21.86 & 1.15 & 1.08 & 0.45 &  --- & A & 10 08 17.93 & +12 21 51.1 \\
V79       & RRc     & 0.417855   & 22.90 & 22.69 & 22.52 & 22.15 & 0.83 & 0.47 &  --- &  --- & C & 10 08 18.55 & +12 11 07.5 \\
V80       & RRc     & 0.3866383  & 22.66 & 22.44 & 22.33 & 22.03 & 0.50 & 0.47 & 0.31 & 0.19 & A & 10 08 18.99 & +12 16 23.3 \\
V81       & RRab    & 0.596522   & 23.15 & 22.82 & 22.50 & 22.23 & 0.85 & 0.62 & 0.59 &  --- & A & 10 08 19.12 & +12 15 31.5 \\
V82       & RRab    & 0.5711366  & 23.12 & 22.77 & 22.72 & 22.14 & 1.34 & 1.02 &  --- &  --- & A & 10 08 19.43 & +12 21 59.7 \\
V83       & RRab    & 0.6778582  & 22.36 & 21.82 & 21.62 & 21.17 & 0.53 & 0.38 &  --- &  --- & C & 10 08 19.84 & +12 18 27.4 \\
V84       & RRab    & 0.615537   & 22.90 & 22.52 & 22.49 & 21.84 & 1.71 & 1.08 & 0.31 &  --- & A & 10 08 19.88 & +12 14 47.4 \\
V85       & RRc     & 0.3175981  & 22.67 & 22.45 & 22.13 & 22.11 & 0.61 & 0.50 & 0.25 &  --- & A & 10 08 19.90 & +12 18 58.7 \\
V86       & RRc     & 0.2851411  & 23.01 & 22.55 & 22.41 &  ---  & 0.92 & 0.84 & 0.29 &  --- & B & 10 08 19.91 & +12 26 28.2 \\
V87       & RRab    & 0.651948   & 23.02 & 22.64 & 22.35 & 21.91 & 0.60 & 0.46 & 0.34 &  --- & A & 10 08 19.93 & +12 15 55.3 \\
V88       & RRc     & 0.3173750  & 22.96 & 22.72 & 22.57 & 22.45 & 0.73 & 0.75 &  --- &  --- & B & 10 08 19.99 & +12 23 03.1 \\
V89       & RRab    & 0.5300555  & 22.85 & 22.64 & 22.49 & 22.27 & 1.40 & 0.92 & 0.83 &  --- & A & 10 08 20.11 & +12 16 12.9 \\
V91       & RRab    & 0.602603   & 23.14 & 22.72 & 22.58 &  ---  & 0.69 & 0.72 & 0.64 &  --- & A & 10 08 20.58 & +12 24 18.3 \\
V92*      & RRab    & 0.6304080  & 22.44 & 22.23 & 21.96 & 21.75 & 0.99 & 0.80 & 0.67 & 0.12 & B & 10 08 20.77 & +12 19 26.9 \\
V93       & RRab    & 0.5911859  & 22.89 & 22.47 & 22.26 & 21.86 & 1.00 & 0.70 & 0.50 & 0.25 & A & 10 08 20.94 & +12 16 32.6 \\
V96       & RRab    & 0.6429762  & 22.68 & 22.43 & 22.22 & 22.04 & 1.08 & 0.76 & 0.70 &  --- & A & 10 08 21.47 & +12 18 54.8 \\
V97       & RRab    & 0.591565   & 23.14 & 22.77 & 22.62 & 22.21 & 0.58 & 0.35 & 0.25 &  --- & C & 10 08 21.52 & +12 23 45.1 \\
V98*      & RRab    & 0.626239   & 22.96 & 22.63 & 22.47 & 22.24 & 1.27 & 0.89 & 0.70 &  --- & A & 10 08 21.53 & +12 14 27.1 \\
V99       & RRab    & 0.574072   & 23.28 & 22.92 & 22.66 & 22.45 & 1.19 & 0.73 &  --- &  --- & B & 10 08 21.68 & +12 15 46.3 \\
V101      & RRab    & 0.7071830  & 22.92 & 22.62 & 22.29 & 22.00 & 0.86 & 0.78 & 0.78 &  --- & B & 10 08 21.86 & +12 21 54.3 \\
V102      & RRab    & 0.952880   & 22.88 & 22.39 & 22.57 & 22.34 & 0.73 & 1.18 &  --- &  --- & C & 10 08 22.21 & +12 15 54.6 \\
V105      & RRab    & 0.5541252  & 23.10 & 22.74 & 22.67 & 22.01 & 1.41 & 1.03 & 0.30 &  --- & A & 10 08 22.92 & +12 14 36.6 \\
V106      & RRab    & 0.6021605  & 22.90 & 22.51 & 22.21 & 22.02 & 1.31 & 0.97 & 0.57 &  --- & A & 10 08 23.12 & +12 16 16.8 \\
V107      & RRab    & 0.5573474  & 22.94 & 22.63 & 22.47 & 22.37 & 1.11 & 1.04 & 0.75 &  --- & B & 10 08 23.20 & +12 16 35.9 \\
V108      & RRab    & 0.5467398  & 22.59 & 22.15 & 21.84 & 21.49 & 0.96 & 0.52 & 0.43 & 0.40 & C & 10 08 23.36 & +12 19 07.1 \\
V110      & RRab    & 0.6124265  & 22.85 & 22.57 & 22.50 & 22.04 & 0.84 & 0.49 & 0.21 &  --- & C & 10 08 23.41 & +12 20 30.7 \\
V111      & RRab    & 0.6183992  & 22.82 & 22.56 & 22.22 & 22.11 & 1.26 & 0.93 & 0.50 &  --- & A & 10 08 23.80 & +12 17 26.0 \\
V112      & RRab    & 0.5880014  & 22.31 & 21.81 & 21.47 & 21.12 & 0.69 & 0.44 &  --- &  --- & C & 10 08 24.26 & +12 17 27.5 \\
V116      & RRab    & 0.5593915  & 22.92 & 22.60 & 22.26 & 22.18 & 1.34 & 1.19 & 0.66 &  --- & B & 10 08 24.72 & +12 17 17.1 \\
V117      & RRab    & 0.4948878  & 22.43 & 22.19 & 22.13 & 21.60 & 0.52 & 0.50 &  --- &  --- & B & 10 08 24.76 & +12 19 13.5 \\
V118      & RRab    & 0.5966387  & 22.89 & 22.61 & 22.35 & 22.19 & 1.06 & 0.75 & 0.70 &  --- & A & 10 08 25.36 & +12 17 31.4 \\
V119      & RRc     & 0.3769470  & 22.95 & 22.66 & 22.52 &  ---  & 0.51 & 0.47 & 0.34 &  --- & A & 10 08 25.55 & +12 23 28.1 \\
V121*     & RRab    & 0.5797797  & 22.94 & 22.69 & 22.33 & 21.92 & 1.24 & 0.85 & 0.47 &  --- & A & 10 08 25.61 & +12 15 35.5 \\
V124      & RRab    & 0.5866781  & 23.08 & 22.72 & 19.95 & 22.16 & 1.21 & 0.82 & 0.73 &  --- & B & 10 08 26.01 & +12 18 49.9 \\
V125      & RRab    & 0.5942164  & 22.96 & 22.76 & 22.58 & 22.28 & 1.04 & 0.78 & 0.58 &  --- & B & 10 08 26.22 & +12 21 53.4 \\
V127      & RRab    & 0.5511360  & 22.18 & 21.64 & 21.22 & 20.88 & 0.55 & 0.29 & 0.20 & 0.10 & B & 10 08 26.42 & +12 21 42.4 \\
V128      & RRab    & 0.5398457  & 23.00 & 22.76 & 22.94 &  ---  & 1.35 & 1.00 & 0.31 &  --- & A & 10 08 26.61 & +12 24 11.1 \\
V132      & RRab    & 0.5881914  & 23.09 & 22.76 & 22.46 & 22.18 & 1.26 & 1.00 & 0.82 &  --- & A & 10 08 27.00 & +12 15 11.7 \\
V134      & RRab    & 0.8547496  & 22.94 & 22.57 & 22.48 & 21.98 & 0.80 & 0.64 &  --- &  --- & A & 10 08 27.26 & +12 15 44.5 \\
V135      & RRab    & 0.5886723  & 23.03 & 22.53 & 22.25 & 21.99 & 1.35 & 0.77 & 0.51 &  --- & A & 10 08 27.35 & +12 20 42.7 \\
V137      & RRab    & 0.5978996  & 23.19 & 22.81 & 22.74 & 22.02 & 1.08 & 0.83 &  --- &  --- & A & 10 08 27.61 & +12 20 12.7 \\
V138      & RRab    & 0.5881765  & 23.13 & 22.79 & 22.75 & 22.32 & 0.84 & 0.69 & 0.42 &  --- & A & 10 08 27.87 & +12 16 14.6 \\
V141      & RRc     & 0.3667905  & 22.80 & 22.49 & 22.28 & 22.10 & 0.54 & 0.43 & 0.46 &  --- & A & 10 08 28.13 & +12 15 29.3 \\
V142*     & RRab    & 0.5871330  & 22.93 & 22.51 & 22.21 & 21.87 & 1.12 & 0.82 & 0.43 &  --- & A & 10 08 28.23 & +12 17 13.3 \\
V148      & RRab    & 0.5599570  & 22.35 & 21.86 & 21.51 & 21.07 & 0.58 & 0.37 & 0.31 & 0.28 & B & 10 08 28.84 & +12 20 12.2 \\
V149      & RRc     & 0.3703344  & 22.99 & 22.67 & 22.45 & 22.11 & 0.65 & 0.47 & 0.36 &  --- & A & 10 08 28.94 & +12 15 01.2 \\
V151      & RRab    & 0.6250526  & 22.80 & 22.46 & 22.15 & 21.91 & 0.69 & 0.59 & 0.56 & 0.44 & C & 10 08 29.40 & +12 18 21.3 \\
V152      & RRab    & 0.5552238  & 23.07 & 22.76 & 22.56 & 22.21 & 1.20 & 0.77 & 1.64 &  --- & A & 10 08 29.40 & +12 16 33.3 \\
V153      & RRab    & 0.5612972  & 23.08 & 22.70 & 22.82 & 22.20 & 1.12 & 1.07 & 0.19 &  --- & A & 10 08 29.41 & +12 23 48.3 \\
V156      & RRab    & 0.646932   & 23.35 & 22.92 & 22.58 & 22.23 & 0.64 & 0.57 & 0.55 & 0.36 & C & 10 08 30.41 & +12 17 16.3 \\
V157      & RRab    & 0.4903282  & 22.94 & 22.80 & 22.44 & 22.47 & 1.45 & 0.93 &  --- &  --- & B & 10 08 30.55 & +12 18 37.8 \\
V158      & RRc     & 0.3446077  & 22.87 & 22.58 & 22.38 & 22.12 & 0.61 & 0.45 &  --- &  --- & B & 10 08 30.60 & +12 15 11.0 \\
V165      & RRc     & 0.3231045  & 23.01 & 22.78 & 22.76 & 22.34 & 0.80 & 0.49 & 0.69 & 0.21 & A & 10 08 31.56 & +12 21 48.1 \\
V166      & RRc     & 0.3872139  & 22.83 & 22.58 & 22.41 & 21.94 & 0.58 & 0.42 & 0.36 & 0.82 & B & 10 08 31.55 & +12 15 59.4 \\
V168*     & RRab    & 0.579400   & 23.00 & 22.64 & 22.57 & 21.96 & 1.12 & 1.02 & 0.52 &  --- & A & 10 08 31.78 & +12 23 35.3 \\
V169      & RRab    & 0.5494407  & 23.25 & 22.85 & 22.68 & 22.19 & 1.16 & 0.91 & 0.35 & 0.30 & A & 10 08 31.94 & +12 17 01.9 \\
V170      & RRc     & 0.3890384  & 22.61 & 22.39 & 22.15 & 21.98 & 0.68 & 0.51 & 0.53 &  --- & B & 10 08 31.94 & +12 16 56.2 \\
V171*     & RRab    & 0.5608915  & 22.98 & 22.76 & 22.12 & 21.81 & 1.13 & 1.09 & 0.70 &  --- & A & 10 08 32.06 & +12 13 49.0 \\
V172      & RRab    & 0.615115   & 23.33 & 22.87 & 22.55 & 22.26 & 0.85 & 0.52 & 0.42 &  --- & C & 10 08 32.13 & +12 16 35.3 \\
V174      & RRab    & 0.5950424  & 23.02 & 22.67 & 22.42 & 22.14 & 0.73 & 0.65 & 0.32 &  --- & B & 10 08 32.24 & +12 16 33.5 \\
V175      & RRab    & 0.6380987  & 22.88 & 22.60 & 22.38 & 22.02 & 0.43 & 0.39 & 0.46 &  --- & C & 10 08 32.51 & +12 16 04.9 \\
V176      & RRab    & 0.573147   & 22.38 & 21.85 & 21.51 & 21.11 & 0.74 & 0.54 &  --- &  --- & C & 10 08 33.07 & +12 17 15.5 \\
V180*     & RRab    & 0.619147   & 22.96 & 22.72 & 22.43 & 22.19 & 1.24 & 0.85 & 0.49 &  --- & A & 10 08 34.12 & +12 23 43.2 \\
V185      & RRab    & 0.738784   & 22.88 & 22.49 & 22.16 & 21.80 & 1.15 & 0.86 & 0.69 &  --- & A & 10 08 35.94 & +12 22 49.2 \\
V186      & RRab    & 0.6128591  & 22.89 & 22.58 & 22.27 & 22.03 & 1.26 & 0.89 & 0.56 &  --- & A & 10 08 36.05 & +12 21 27.8 \\
V187      & RRab    & 0.631086   & 22.80 & 22.43 & 22.18 & 21.81 & 0.70 & 0.58 & 0.36 &  --- & A & 10 08 36.09 & +12 19 33.3 \\
V188      & RRab    & 0.622437   & 23.20 & 22.74 & 22.65 & 22.07 & 0.42 & 0.50 & 0.21 &  --- & B & 10 08 36.31 & +12 27 15.9 \\
V189      & RRab    & 0.6313365  & 23.01 & 22.57 & 22.17 & 22.01 & 1.22 & 0.93 & 0.73 &  --- & A & 10 08 36.88 & +12 19 45.0 \\
V192      & RRab    & 0.667686   & 22.93 & 22.59 & 22.38 & 22.10 & 0.60 & 0.66 & 0.60 &  --- & A & 10 08 37.44 & +12 20 44.6 \\
V196      & RRc     & 0.352561   & 22.86 & 22.65 & 22.55 & 22.16 & 0.80 & 0.61 & 0.37 &  --- & C & 10 08 38.87 & +12 21 52.8 \\
V198      & RRc     & 0.364983   & 22.95 & 22.68 & 22.52 & 22.17 & 0.56 & 0.41 & 0.38 &  --- & A & 10 08 39.35 & +12 16 09.2 \\
V201*     & RRab    & 0.5663093  & 23.05 & 22.76 & 22.58 & 22.04 & 1.26 & 1.02 & 0.42 &  --- & A & 10 08 40.00 & +12 19 18.4 \\
V202*     & RRab    & 0.581068   & 23.05 & 22.69 & 22.58 & 21.95 & 1.23 & 0.67 & 0.21 &  --- & A & 10 08 40.10 & +12 14 28.5 \\
V204      & RRab    & 0.748624   & 22.86 & 22.54 & 22.18 & 21.98 & 1.55 & 1.01 & 0.79 & 0.65 & A & 10 08 40.36 & +12 18 53.7 \\
V205*     & RRab    & 0.6199640  & 22.92 & 22.60 & 22.29 & 22.16 & 1.08 & 0.86 & 0.48 &  --- & A & 10 08 40.63 & +12 19 18.7 \\
V207      & RRab    & 0.568369   & 22.98 & 22.66 & 22.56 & 22.29 & 1.25 & 0.86 & 0.43 &  --- & B & 10 08 41.00 & +12 19 57.9 \\
V213      & RRab    & 0.602289   & 22.75 & 22.53 & 22.48 & 21.96 & 1.22 & 1.00 & 0.80 &  --- & A & 10 08 42.91 & +12 20 38.9 \\
V214*     & RRc     & 0.3190927  & 22.83 & 22.60 & 22.65 & 22.07 & 0.72 & 0.63 & 0.25 &  --- & A & 10 08 44.37 & +12 21 30.6 \\
V216      & RRab    & 0.723219   & 22.98 & 22.64 & 22.26 & 21.99 & 0.79 & 0.71 & 0.30 & 0.69 & B & 10 08 44.45 & +12 21 33.4 \\
V217      & RRab    & 0.541640   & 23.04 & 22.72 & 22.49 & 22.25 & 0.88 & 1.01 & 0.52 & 0.66 & C & 10 08 45.15 & +12 20 00.5 \\
V218      & RRc     & 0.371302   & 22.98 & 22.68 & 22.49 & 22.09 & 0.58 & 0.40 & 0.35 &  --- & B & 10 08 47.40 & +12 21 08.0 \\
V219*     & RRab    & 0.587327   & 23.12 & 22.77 & 22.47 & 22.09 & 1.53 & 0.90 & 0.75 &  --- & A & 10 08 48.15 & +12 16 28.4 \\
V220      & RRc     & 0.3124520  & 22.94 & 22.69 & 22.73 & 22.29 & 0.66 & 0.58 & 0.31 &  --- & B & 10 08 48.32 & +12 19 24.6 \\
V221*     & RRab    & 0.665057   & 22.76 & 22.50 & 22.10 & 21.72 & 1.20 & 1.02 & 0.64 &  --- & C & 10 08 48.38 & +12 26 51.8 \\
V222*     & RRab    & 0.566679   & 23.06 & 22.77 & 22.46 & 22.27 & 1.05 & 0.95 & 0.60 &  --- & A & 10 08 48.43 & +12 17 58.0 \\
V223      & RRab    & 0.562598   & 22.91 & 22.72 & 22.55 & 22.35 & 0.86 & 0.74 & 0.38 &  --- & B & 10 08 48.54 & +12 17 52.5 \\
V224      & RRab    & 0.647379   & 23.07 & 22.68 & 22.48 & 21.97 & 0.52 & 0.33 & 0.38 &  --- & B & 10 08 48.68 & +12 18 47.6 \\
V225      & RRc     & 0.379586   & 23.01 & 22.70 & 22.40 & 22.15 & 0.54 & 0.43 & 0.40 &  --- & A & 10 08 48.95 & +12 21 41.2 \\
V226      & RRc     & 0.2637345  & 22.76 & 22.50 & 22.55 & 22.15 & 0.58 & 0.50 & 0.24 &  --- & B & 10 08 49.17 & +12 21 26.3 \\
V227      & RRab    & 0.644914   & 23.07 & 22.64 & 22.42 & 22.04 & 0.64 & 0.44 & 0.50 &  --- & C & 10 08 49.36 & +12 14 52.0 \\
V228*     & RRab    & 0.600145   & 23.10 & 22.74 & 22.46 & 22.08 & 1.00 & 0.83 & 0.34 &  --- & A & 10 08 50.71 & +12 18 00.0 \\
V229*     & RRab    & 0.5929263  & 22.92 & 22.71 & 22.74 & 22.31 & 1.49 & 1.02 & 0.41 &  --- & A & 10 08 51.38 & +12 21 54.3 \\
V230*     & RRab    & 0.565908   & 22.98 & 22.67 & 22.42 & 22.21 & 1.06 & 0.81 &  --- &  --- & B & 10 08 51.67 & +12 15 56.6 \\
V231*     & RRab    & 0.623239   & 23.04 & 22.59 & 22.46 & 21.90 & 0.65 & 0.65 & 0.34 &  --- & B & 10 08 51.73 & +12 29 06.5 \\
V232*     & RRab    & 0.581134   & 22.98 & 22.66 & 22.33 & 21.90 & 1.12 & 0.90 & 0.43 &  --- & A & 10 08 52.03 & +12 21 00.5 \\
V234*     & RRab    & 0.719480   & 22.96 & 22.56 & 22.46 & 21.82 & 1.08 & 0.97 & 0.52 &  --- & A & 10 08 52.99 & +12 21 57.9 \\
V235*     & RRab    & 0.6603587  & 22.84 & 22.48 & 22.16 & 21.80 & 1.98 & 1.07 & 0.76 &  --- & A & 10 08 53.20 & +12 15 39.3 \\
V236*     & RRab    & 0.5605958  & 22.99 & 22.62 & 22.77 & 22.02 & 1.03 & 1.04 & 0.35 &  --- & A & 10 08 54.82 & +12 22 16.3 \\
V237*     & RRab    & 0.592160   & 22.86 & 22.51 & 22.38 & 21.94 & 1.22 & 1.23 & 0.42 &  --- & A & 10 08 55.31 & +12 17 41.1 \\
V239*     & RRab    & 0.6390675  & 22.85 & 22.60 & 22.23 & 22.28 & 1.01 & 0.77 & 0.83 &  --- & A & 10 08 55.60 & +12 21 25.1 \\
V240*     & RRab    & 0.601778   & 23.03 & 22.74 & 22.49 & 22.37 & 0.99 & 1.02 &  --- &  --- & C & 10 08 55.82 & +12 14 50.7 \\
V243*     & RRc     & 0.358956   & 23.02 & 22.64 & 22.49 & 22.07 & 0.47 & 0.46 & 0.45 &  --- & B & 10 08 57.04 & +12 13 02.4 \\
V244*     & RRc     & 0.3881747  & 22.97 & 22.67 & 22.53 & 22.20 & 0.59 & 0.50 & 0.43 &  --- & B & 10 08 58.23 & +12 24 54.3 \\
V245*     & RRab    & 0.541998   & 23.16 & 22.65 & 22.44 & 22.18 & 0.36 & 0.79 & 0.54 &  --- & A & 10 08 59.99 & +12 19 49.4 \\
V246*     & RRab    & 0.5475991  & 23.07 & 22.66 & 22.27 & 22.20 & 1.06 & 0.90 &  --- &  --- & B & 10 09 01.43 & +12 12 07.4 \\
V247*     & RRab    & 0.68277    & 22.86 & 22.46 & 22.51 & 21.91 & 1.79 & 1.14 &  --- &  --- & B & 10 09 04.52 & +12 20 41.4 \\
V249*     & RRab    & 0.59676    & 22.96 & 22.67 & 22.62 & 22.07 & 0.75 & 0.69 &  --- &  --- & B & 10 09 06.66 & +12 21 17.7 \\
V250*     & RRab    & 0.58273    & 22.94 & 22.70 & 22.77 & 22.28 & 0.64 & 0.59 &  --- &  --- & C & 10 09 08.17 & +12 24 32.1 \\
V251*     & RRab    & 0.574      & 23.22 & 22.72 & 22.20 & 22.22 & 1.42 & 1.07 &  --- &  --- & C & 10 09 09.47 & +12 19 26.3 \\
\enddata

%\\
%\hline
%\hline
\tablecomments{* indicates suspected Blazhko or double-mode pulsator.}
\end{deluxetable}

%%%%%%%%%%%%%%%%%%%%%%%%%%%%%%%%%%%%%%%%%%%%%%%%%%%%%%%%%%%%%%%%%%%%%%%%%%
%TAB 3
%%%%%%%%%%%%%%%%%%%%%%%%%%%%%%%%%%%%%%%%%%%%%%%%%%%%%%%%%%%%%%%%%%%%%%%%%%

\begin{deluxetable}{ll cccc cccc ccccc}
\centering
\tabletypesize{\tiny}
\tablewidth{0pt}
\tablenum{3}
\tablecaption{AC/spC variable candidates: periods, photometric parameters, and positions.}
\label{tabac}
\tablehead{
\colhead{ID}    &
 \colhead{Period} &
 \colhead{ $\left<B\right>$} &
 \colhead{ $\left<V\right>$} &
 \colhead{ $\left<R\right>$} &
 \colhead{ $\left<I\right>$} & 
 \colhead{$A_B$}  & 
 \colhead{$A_V$}  & 
 \colhead{$A_R$}  & 
 \colhead{$A_I$}  & 
 \colhead{Q}   &
 \colhead{Mode}   &
 \colhead{Mass}   &
 \colhead{RA} &
 \colhead{Dec} 
\\
	\colhead{}&
	\colhead{d}&
	\colhead{mag}&
	\colhead{mag}&
	\colhead{mag}&
	\colhead{mag}&
	\colhead{mag}&
	\colhead{mag}&
	\colhead{mag}&
	\colhead{mag}&&&
	\colhead{M$_{\odot}$}&
	\colhead{hh mm ss}&
	\colhead{dd mm ss}
\\
}
\startdata
V20       & 1.332426  & 21.48 & 20.99 & 20.85 & 20.68 & 1.42 & 1.26 & 0.91 & 0.80 & A & FO & 1.9 & 10 07 59.30 & +12 17 45.5 \\
V27       & 1.6104244 & 20.88 & 20.51 & 20.35 & 19.87 & 1.84 & 1.46 & 0.46 & 0.60 & A & FO & 1.5 & 10 08 05.42 & +12 18 22.9 \\
V34       & 1.660596  & 20.78 & 20.54 & 20.18 & 19.93 & 1.74 & 1.30 & 1.00 & 0.83 & A & FO & 0.8 & 10 08 08.16 & +12 16 57.0 \\
V39       & 1.339882  & 21.12 & 20.75 & 20.32 & 20.02 & 0.87 & 0.68 & 0.37 & 0.20 & A & FU & 2.6 & 10 08 08.89 & +12 19 47.3 \\
V40       & 2.08521   & 20.91 & 20.52 & 20.36 & 19.96 & 0.84 & 0.62 & 0.60 & 0.53 & A & FO & 1.1 & 10 08 09.06 & +12 16 04.0 \\
V41       & 0.6001349 & 21.57 & 21.25 & 21.08 & 20.67 & 0.80 & 0.66 & 0.64 & 0.52 & A & FO & 2.3 & 10 08 09.33 & +12 15 10.5 \\
V45       & 0.8042814 & 21.27 & 20.92 & 20.77 & 20.49 & 0.77 & 0.68 & 0.70 & 0.46 & A & FU & 4.1 & 10 08 11.07 & +12 19 53.9 \\
V62       & 0.8356451 & 21.15 & 20.93 & 20.74 & 20.48 & 0.92 & 0.71 & 0.56 & 0.54 & A & FO & 1.3 & 10 08 15.11 & +12 20 41.4 \\
V69       & 1.656557  & 21.12 & 20.80 & 20.62 & 20.22 & 1.64 & 1.27 & 0.70 & 0.85 & A & FU & 1.4 & 10 08 16.41 & +12 16 51.1 \\
V71       & 0.7634400 & 21.72 & 21.38 & 21.25 & 20.83 & 0.84 & 0.66 & 0.56 & 0.47 & A & FO & 1.5 & 10 08 16.75 & +12 19 19.3 \\
V75       & 0.5443509 & 21.32 & 21.07 & 20.88 & 20.70 & 0.81 & 0.63 & 0.53 & 0.28 & A & FO & 2.5 & 10 08 17.47 & +12 18 04.4 \\
V78       & 1.304405  & 21.18 & 20.75 & 20.47 & 20.21 & 1.30 & 1.05 & 0.79 & 0.23 & A & FO & 2.1 & 10 08 18.15 & +12 18 08.0 \\
V94       & 0.7377763 & 20.90 & 20.62 & 20.46 & 20.05 & 0.91 & 0.68 & 0.26 & 0.26 & A & FO & 3.1 & 10 08 21.12 & +12 17 47.4 \\
V103      & 1.4382564 & 20.82 & 20.48 & 20.23 & 19.87 & 1.76 & 1.36 & 9.99 & 1.00 & A & FO & 1.7 & 10 08 22.29 & +12 18 01.9 \\
V113      & 0.7554243 & 21.18 & 20.89 & 20.67 & 20.39 & 0.89 & 0.60 & 0.37 & 0.38 & A & FO & 2.2 & 10 08 24.27 & +12 20 05.5 \\
V114=HW19 & 1.1232940 & 21.21 & 20.96 & 20.51 & 20.43 & 1.67 & 1.37 & 0.83 & 0.61 & A & FO & 0.9 & 10 08 24.49 & +12 15 04.8 \\
V115      & 1.0984368 & 21.40 & 21.01 & 20.76 & 20.44 & 1.47 & 1.20 & 1.12 & 0.87 & A & FO & 1.6 & 10 08 24.53 & +12 18 23.7 \\
V120      & 0.984594  & 21.13 & 21.00 & 20.82 & 20.46 & 1.46 & 1.21 & 0.67 & 0.53 & A & FO & 0.6 & 10 08 25.58 & +12 16 51.5 \\
V122      & 1.306483  & 21.28 & 20.86 & 20.67 & 20.31 & 1.24 & 1.00 & 0.55 & 0.46 & A & FO & 1.7 & 10 08 25.92 & +12 18 03.3 \\
V123      & 0.754381  & 21.42 & 21.03 & 20.74 & 20.44 & 0.44 & 0.31 & 0.31 & 0.31 & A & FU & 4.8 & 10 08 25.99 & +12 19 46.8 \\
V129=HW23$^*$ & 1.1947745 & 21.04 & 20.72 & 20.44 & 20.18 & 1.40 & 1.08 & 0.68 & 0.30 & A & FU & 2.5 & 10 08 26.62 & +12 18 11.1 \\
V130      & 0.6902270 & 21.39 & 21.09 & 20.99 & 20.60 & 0.70 & 0.62 & 0.69 & 0.52 & A & FO & 2.1 & 10 08 26.85 & +12 18 52.7 \\
V131      & 1.1634407 & 21.27 & 20.92 & 20.65 & 20.39 & 1.39 & 0.99 & 0.50 & 0.41 & A & FU & 2.2 & 10 08 26.86 & +12 19 27.4 \\
V133      & 0.5180820 & 21.32 & 20.90 & 20.63 & 20.30 & 0.63 & 0.46 & 0.25 & 0.15 & A & unc & --- & 10 08 27.02 & +12 18 07.8 \\
V140      & 1.5931487 & 21.16 & 20.71 & 20.39 & 20.09 & 1.27 & 0.94 & 0.66 & 0.36 & A & FO & 1.7 & 10 08 28.08 & +12 18 16.3 \\
V144=HW2  & 1.0098826 & 21.24 & 20.86 & 20.53 & 20.24 & 1.43 & 1.22 & 1.10 & 0.68 & A & FO & 2.2 & 10 08 28.54 & +12 19 23.7 \\
V145      & 1.6100353 & 20.87 & 20.54 & 20.23 & 19.99 & 1.71 & 1.33 & 0.88 & 0.77 & A & FU & 2.0 & 10 08 28.54 & +12 22 23.2 \\
V146=HW16 & 1.4985041 & 21.25 & 20.82 & 19.98 & 20.27 & 1.60 & 1.08 & 1.00 & 0.53 & A & FO & 1.5 & 10 08 28.77 & +12 14 59.9 \\
V147      & 0.6692048 & 21.61 & 21.20 & 20.96 & 20.72 & 0.57 & 0.60 & 0.25 & 0.20 & A & FO & 3.1 & 10 08 28.82 & +12 14 55.5 \\
V150      & 1.3694463 & 21.30 & 20.87 & 20.48 & 20.26 & 0.82 & 0.65 & 0.38 & 0.27 & A & FO & 1.6 & 10 08 29.26 & +12 17 27.5 \\
V154=HW8  & 3.121384  & 20.35 & 19.93 & 19.52 & 19.31 & 1.11 & 0.88 & 0.45 & 0.47 & A & FU & 2.3 & 10 08 30.04 & +12 16 13.3 \\
V155      & 0.9951689 & 20.80 & 20.52 & 20.40 & 20.09 & 0.77 & 0.67 & 0.46 & 0.31 & A & FO & 2.2 & 10 08 30.22 & +12 17 53.5 \\
V159      & 1.706686  & 21.26 & 20.88 & 20.64 & 20.36 & 1.39 & 0.98 & 0.77 & 0.37 & A & FU & 1.5 & 10 08 30.63 & +12 21 54.1 \\
V160      & 1.2405284 & 20.91 & 20.62 & 20.51 & 20.09 & 1.63 & 1.40 & 0.53 & 0.68 & A & FU & 2.3 & 10 08 30.67 & +12 19 19.4 \\
V163      & 1.2032229 & 21.33 & 20.91 & 20.77 & 20.39 & 1.40 & 1.15 & 0.83 & 0.77 & A & FO & 1.8 & 10 08 31.09 & +12 18 11.1 \\
V164=HW17 & 2.043527  & 20.63 & 20.26 & 20.03 & 19.74 & 1.45 & 1.03 & 0.95 & 0.43 & A & FU & 2.4 & 10 08 31.39 & +12 16 54.5 \\
V173      & 0.7528587 & 21.00 & 20.71 & 20.58 & 20.28 & 0.74 & 0.62 & 0.44 & 0.49 & A & FO & 2.8 & 10 08 32.23 & +12 18 46.3 \\
V177      & 0.6462353 & 21.37 & 21.06 & 20.89 & 20.54 & 0.80 & 0.56 & 0.42 & 0.31 & A & FO & 2.5 & 10 08 33.59 & +12 18 14.3 \\
V179=HW3  & 0.9871234 & 21.33 & 21.03 & 9.999 & 20.47 & 1.37 & 1.10 & 9.99 & 0.73 & A & FU & 2.2 & 10 08 34.10 & +12 18 43.9 \\
V181      & 1.3872086 & 21.00 & 20.55 & 20.27 & 19.88 & 1.28 & 1.02 & 0.35 & 0.51 & A & FO & 2.6 & 10 08 34.56 & +12 15 57.6 \\
V182=HW10 & 1.494313  & 21.21 & 20.86 & 20.62 & 20.30 & 1.57 & 1.21 & 0.57 & 0.55 & A & FU & 1.6 & 10 08 34.67 & +12 15 47.0 \\
V184      & 0.7579201 & 21.01 & 20.67 & 20.51 & 20.14 & 0.91 & 0.68 & 0.48 & 0.29 & A & FO & 3.7 & 10 08 35.17 & +12 18 56.6 \\
V193      & 0.7075682 & 21.14 & 20.82 & 20.61 & 20.31 & 0.78 & 0.60 & 0.51 & 0.31 & A & FO & 3.1 & 10 08 37.61 & +12 16 42.5 \\
V194      & 1.0362154 & 21.57 & 21.10 & 20.79 & 20.54 & 1.92 & 0.67 & 0.46 & 0.33 & A & FO & 2.3 & 10 08 37.95 & +12 18 28.6 \\
V195      & 1.222193  & 20.90 & 20.63 & 20.22 & 20.20 & 1.53 & 1.18 & 0.84 & 0.39 & A & FO & 1.3 & 10 08 38.22 & +12 19 54.2 \\
V197      & 2.449166  & 20.70 & 20.24 & 19.92 & 19.60 & 0.62 & 0.46 & 0.31 & 0.24 & A & FU & 2.7 & 10 08 39.34 & +12 18 00.0 \\
V199=HW1  & 1.493284  & 20.77 & 20.49 & 20.58 & 20.09 & 2.01 & 1.49 & 0.30 & 0.54 & A & FU & 2.0 & 10 08 39.39 & +12 22 11.9 \\
V203      & 0.624501  & 21.05 & 20.61 & 20.27 & 19.92 & 0.44 & 0.35 & 0.33 & 0.23 & A & unc & --- & 10 08 40.15 & +12 18 10.5 \\
V208      & 0.7337240 & 20.97 & 20.66 & 20.53 & 20.25 & 0.83 & 0.72 & 0.45 & 0.40 & A & FO & 3.4 & 10 08 41.33 & +12 19 09.0 \\
V209      & 1.202899  & 21.73 & 21.29 & 21.03 & 20.63 & 0.89 & 0.69 & 0.34 & 0.41 & A & FO & 1.2 & 10 08 41.74 & +12 17 39.8 \\
V211      & 1.353783  & 21.27 & 20.86 & 20.79 & 20.39 & 1.31 & 1.22 & 0.40 & 0.42 & A & FO & 1.6 & 10 08 42.53 & +12 14 42.2 \\
V212      & 0.967586  & 21.20 & 20.87 & 20.55 & 20.32 & 0.68 & 0.35 & 0.31 & 0.56 & C & FO & 1.8 & 10 08 42.79 & +12 13 04.2 \\
V215      & 1.093596  & 21.37 & 21.12 & 21.17 & 20.58 & 1.70 & 1.11 & 0.90 & 0.24 & A & FO & 0.8 & 10 08 44.41 & +12 16 02.5 \\
V241      & 0.98188   & 21.78 & 20.98 & 20.62 & 20.28 & 0.29 & 0.25 & 0.31 & 0.33 & C & unc & --- & 10 08 56.20 & +12 35 07.8 \\
V248      & 0.907955  & 21.11 & 20.89 & 20.83 & 20.43 & 0.80 & 0.67 & 0.26 & 0.37 & A & FO & 1.2 & 10 09 04.79 & +12 21 10.7

\enddata 
\tablecomments{* indicates suspected Blazhko or double-mode pulsator.}
\end{deluxetable}

%%%%%%%%%%%%%%%%%%%%%%%%%%%%%%%%%%%%%%%%%%%%%%%%%%%%%%%%%%%%%%%%%%%%%%%%%%
%TAB 4
%%%%%%%%%%%%%%%%%%%%%%%%%%%%%%%%%%%%%%%%%%%%%%%%%%%%%%%%%%%%%%%%%%%%%%%%%%
\begin{deluxetable}{ll cccc cccc cc}
\centering
\tabletypesize{\tiny}
\tablewidth{0pt}
\tablenum{4}
\tablecaption{Positions and photometry for the candidate LPV stars.}
%\tablenotemark{*}. 
%}
\label{tablpv}
\tablehead{
\colhead{ID}    &
 \colhead{``period''} &
 \colhead{ $\left<B\right>$} &
 \colhead{ $\left<V\right>$} &
 \colhead{ $\left<R\right>$} &
 \colhead{ $\left<I\right>$} & 
 \colhead{$A_B$}  & 
 \colhead{$A_V$}  & 
 \colhead{$A_R$}  & 
 \colhead{$A_I$}  & 
 \colhead{RA} &
 \colhead{Dec}
\\
	\colhead{}&
	\colhead{d}&
	\colhead{mag}&
	\colhead{mag}&
	\colhead{mag}&
	\colhead{mag}&
	\colhead{mag}&
	\colhead{mag}&
	\colhead{mag}&
	\colhead{mag}&
	\colhead{hh mm ss}&
	\colhead{dd mm ss}
\\
}
\startdata
V29  & 137.33  & 20.73 & 19.17 & 18.25 & 17.39 & 0.87 & 0.78 & 0.78 & 0.65 & 10 08 05.93 & +12 15 22.7 \\
V30  & 22.966  & 21.06 & 19.59 & 18.77 & 18.07 & 0.16 & 0.12 & 0.12 & 0.11 & 10 08 06.43 & +12 16 26.9 \\
V33  & 13.515  & 20.98 & 19.28 & 18.36 & 17.56 & 0.13 & 0.20 & 0.13 & 0.13 & 10 08 08.03 & +12 15 18.1 \\
V50  & 52.523  & 20.83 & 19.31 & 18.51 & 17.79 & 0.38 & 0.26 & 0.33 & 0.21 & 10 08 12.29 & +12 23 19.7 \\
V61  & 13.5399 & 20.91 & 19.11 & 18.10 & 17.34 & 0.26 & 0.20 & 0.17 & 0.11 & 10 08 14.59 & +12 18 01.9 \\
V90  & 24.226  & 20.93 & 19.49 & 18.68 & 17.93 & 0.22 & 0.16 & 0.10 & 0.03 & 10 08 20.47 & +12 21 04.0 \\
V95  & 29.027  & 20.95 & 19.11 & 18.11 & 17.33 & 0.53 & 0.23 & 0.18 & 0.11 & 10 08 21.38 & +12 14 42.9 \\
V100 & 746     & 21.75 & 19.63 & 18.32 & 17.46 & 0.60 & 1.02 & 0.54 & 0.64 & 10 08 21.77 & +12 17 25.1 \\
V126 & 451.9   & 21.55 & 19.45 & 18.40 & 17.46 & 0.47 & 0.31 & 0.43 & 0.17 & 10 08 26.36 & +12 14 02.8 \\
V136 & 35.828  & 21.61 & 19.98 & 19.25 & 17.75 & 0.88 & 1.39 & 1.10 & 1.01 & 10 08 27.52 & +12 16 54.0 \\
V139 & 86.00   & 20.90 & 19.53 & 18.87 & 18.05 & 0.29 & 0.16 & 0.40 & 0.13 & 10 08 27.95 & +12 17 56.5 \\
V143 & 20.170  & 20.77 & 19.43 & 18.48 & 17.60 & 0.48 & 0.26 & 0.24 & 0.22 & 10 08 28.51 & +12 19 48.8 \\
V162 & 65.075  & 20.72 & 18.80 & 17.84 & 17.03 & 0.25 & 0.21 & 0.21 & 0.03 & 10 08 31.04 & +12 17 07.8 \\
V167 & 63.334  & 20.97 & 19.70 & 18.84 & 17.66 & 0.65 & 0.69 & 0.92 & 0.27 & 10 08 31.76 & +12 18 18.1 \\
V178 & 23.708  & 21.27 & 19.98 & 19.18 & 18.58 & 0.16 & 0.14 & 0.02 & 0.06 & 10 08 34.06 & +12 23 16.7 \\
V183 & 73.73   & 21.43 & 19.26 & 18.14 & 17.20 & 0.44 & 0.43 & 0.40 & 0.24 & 10 08 34.72 & +12 18 37.7 \\
V190 & 16.6496 & 20.70 & 19.16 & 18.35 & 17.57 & 0.23 & 0.17 & 0.15 & 0.12 & 10 08 36.92 & +12 20 11.5 \\
V200 & 30.594  & 21.93 & 19.80 & 18.22 & 17.35 & 0.38 & 0.76 & 0.63 & 0.38 & 10 08 39.90 & +12 22 14.8 \\
V238 & 36.707  & 21.06 & 19.70 & 18.82 & 18.08 & 0.38 & 0.38 & 0.34 & 0.40 & 10 08 55.53 & +12 14 01.7

\enddata 
\end{deluxetable}
\newpage

%%%%%%%%%%%%%%%%%%%%%%%%%%%%%%%%%%%%%%%%%%%%%%%%%%%%%%%%%%%%%%%%%%%%%%%%%%
%TAB 5
%%%%%%%%%%%%%%%%%%%%%%%%%%%%%%%%%%%%%%%%%%%%%%%%%%%%%%%%%%%%%%%%%%%%%%%%%%

\begin{deluxetable}{l cccc cc}
\centering
\tabletypesize{\tiny}
\tablewidth{0pt}
\tablenum{5}
\tablecaption{Suspected variables of unknown type}
\label{tabcan}
\tablehead{
\colhead{ID}    &
 \colhead{ $\left<B\right>$} &
 \colhead{ $\left<V\right>$} &
 \colhead{ $\left<R\right>$} &
 \colhead{ $\left<I\right>$} & 
 \colhead{RA} &
 \colhead{Dec}
\\
	\colhead{}&
	\colhead{mag}&
	\colhead{mag}&
	\colhead{mag}&
	\colhead{mag}&
	\colhead{hh mm ss}&
	\colhead{dd mm ss}
\\
}
\startdata
V12  & 23.23 & 22.46 & 22.02 & 21.88 & 10 07 49.83 & +12 34 06.3 \\
V46  & 22.84 & 22.77 & 22.48 & 22.31 & 10 08 11.32 & +12 16 26.8 \\
V48  & 23.02 & 22.52 & 21.85 & 23.02 & 10 08 12.13 & +12 17 19.0 \\
V54  & 23.10 & 22.72 & 22.68 & 22.28 & 10 08 12.54 & +12 15 55.9 \\
V55  & 22.87 & 22.75 & 22.45 & 22.41 & 10 08 13.06 & +12 10 48.4 \\
V104 & 23.10 & 22.84 & 23.34 & 22.81 & 10 08 22.34 & +12 16 26.6 \\
V109 & 22.71 & 22.73 & 22.44 & 22.10 & 10 08 23.38 & +12 12 50.6 \\
V161 & 24.06 & 23.48 & 23.00 & 22.93 & 10 08 30.88 & +12 17 17.1 \\
V191 & 23.46 & 22.93 & 22.29 & 22.32 & 10 08 37.37 & +12 13 01.2 \\
V206 & 22.07 & 21.91 & 21.79 & 21.37 & 10 08 40.90 & +12 20 58.4 \\
V210 & 22.91 & 22.81 & 22.17 & 22.17 & 10 08 42.23 & +12 16 10.8 \\
V233 & 22.74 & 21.68 & 21.54 & 21.15 & 10 08 52.16 & +12 23 49.2 \\
V242 & 21.59 & 21.37 & 21.47 & 20.74 & 10 08 56.21 & +12 20 02.7 \\

\enddata 
\end{deluxetable}

\newpage

%%%%%%%%%%%%%%%%%%%%%%%%%%%%%%%%%%%%%%%%%%%%%%%%%%%%%%%%%%%%%%%%%%%%%%%%%%
%TAB 6
%%%%%%%%%%%%%%%%%%%%%%%%%%%%%%%%%%%%%%%%%%%%%%%%%%%%%%%%%%%%%%%%%%%%%%%%%%

\begin{deluxetable}{lrrcccrr}
\tabletypesize{\tiny}
\tablewidth{0pt}
\tablenum{6}
\tablecaption{Mean RRL properties of selected dwarf Spheroidal galaxies.}
%\tablenotemark{*}. 
%}
%\scriptsize{tiny}
\label{tabmean}
\tablehead{
\colhead{Galaxy's name}    &
\colhead{$\left<\hbox{\rm Pab}\right>$}     &
\colhead{$\left<\hbox{\rm Pc}\right>$}      &
 \colhead{N$_{RRab}$}  &
 \colhead{N$_{TOT}$} &
 \colhead{N$_{RRc}$/N$_{TOT}$} &
 \colhead{M/M$_{\odot}$} &
 \colhead{$<$[Fe/H]$>$} 
\\
	\colhead{}&
	\colhead{d}&
	\colhead{d}&
	\colhead{}&
	\colhead{}&
	\colhead{}&
	\colhead{}&
	\colhead{dex}
}
\startdata
UFDs      & 0.640  $\pm$ 0.010  ($\sigma$= 0.06)  & 0.377 $\pm$ 0.008 ($\sigma$= 0.03)  & 43     &  61    &  0.29 &  $\le 0.1\times10^6$    &$\le$--2.0\\       
Draco    & 0.615 $\pm$ 0.003 ($\sigma$= 0.04)  & 0.389 $\pm$ 0.004 ($\sigma$= 0.03)  & 214    &  270   &  0.21 &  $0.3\times10^6$   &--1.93\\       
Carina   & 0.635 $\pm$ 0.006 ($\sigma$= 0.05)  & 0.370  $\pm$ 0.010  ($\sigma$= 0.05)  & 63     &  82    &  0.23 &  $0.4\times10^6$   &--1.72\\       
Tucana   & 0.604 $\pm$ 0.004 ($\sigma$= 0.06)  & 0.361 $\pm$ 0.005 ($\sigma$= 0.05)  & 216    &  298   &  0.27 &  $0.6\times10^6$   &--1.95\\       
Sculptor & 0.590 $\pm$ 0.007 ($\sigma$= 0.08)  & 0.337 $\pm$ 0.004 ($\sigma$= 0.04)  & 131    &  221   &  0.41 &  $2.3\times10^6$   &--1.68\\       
Cetus    & 0.614 $\pm$ 0.002 ($\sigma$= 0.04)  & 0.383 $\pm$ 0.004 ($\sigma$= 0.04)  & 503    &  630   &  0.20 &  $2.6\times10^6$   &--1.90\\       
Leo~I    & 0.596 $\pm$ 0.001 ($\sigma$= 0.05)  & 0.357  $\pm$ 0.002  ($\sigma$= 0.03)  & 81     &  95    &  0.15 &  $5.5\times10^6$   &--1.43       
\enddata 
\tablecomments{N$_{TOT}$ does not include d--type RRLs. Mass of the
  galaxy is given assuming a stellar mass-to-light ratio of 1 \citep{mcconnachie12}. 
}
\end{deluxetable}

%%%%%%%%%%%%%%%%%%%%%%%%%%%%%%%%%%%%%%%%%%%%%%%%%%%%%%%%%%%%%%%%%%%%%%%%%%
%TAB 7
%%%%%%%%%%%%%%%%%%%%%%%%%%%%%%%%%%%%%%%%%%%%%%%%%%%%%%%%%%%%%%%%%%%%%%%%%%

\begin{deluxetable}{lccc}
\centering
\tabletypesize{\tiny}
\tablewidth{0pt}
\tablenum{7}
\tablecaption{Mean properties for RR Lyraes in the Galactic halo, globular clusters, classical and UFD dwarf galaxies.}
%\tablenotemark{*}. 
%}
%\scriptsize{tiny}
\label{tabcomp}
\tablehead{
\colhead{Galaxy's name}    &
\colhead{$\left<\hbox{\rm Pab}\right>$}     &
 \colhead{N$_{RRab}$}  &
 \colhead{N$_{RR_{OoI}}$/N$_{RRab}$}
\\
	\colhead{}&
	\colhead{d}&
	\colhead{}&
	\colhead{}
}
\startdata
Inner halo a     & 0.586 $\pm$ 0.002 ($\sigma$= 0.08)   & 1898      &  0.64\\       
Inner halo b     & 0.583 $\pm$ 0.001 ($\sigma$= 0.08)   & 4064      &  0.67\\       
Outer halo a     & 0.577 $\pm$ 0.001 ($\sigma$= 0.07)   & 5659      &  0.73\\       
Outer halo b     & 0.580 $\pm$ 0.001 ($\sigma$= 0.07)   & 2268      &  0.69\\       
GCs$^a$          & 0.593 $\pm$ 0.004 ($\sigma$= 0.12)   & 1054       &  0.67\\       
dSphs \& UFDs     & 0.610 $\pm$ 0.001 ($\sigma$= 0.05)   & 1299       &  0.78      
\enddata 
\tablecomments{
$^a$ The GCs included here are  \ngc{1851}, \ngc{1904}, \ngc{3201}, \ngc{362},
\ngc{4147}, \ngc{4590}, \ngc{4833}, \ngc{5024}, \ngc{5139},
 \ngc{5272}, \ngc{5286}, \ngc{5466}, \ngc{5897}, \ngc{5904}, \ngc{5986}, \ngc{6101}, \ngc{6121},
 \ngc{6171}, \ngc{6205}, \ngc{6229}, \ngc{6266}, \ngc{6333}, \ngc{6341}, \ngc{6362}, \ngc{6626}, \ngc{6715}
 \ngc{6809}, \ngc{6864}, \ngc{6934}, \ngc{6981}, \ngc{7006}, \ngc{7089}, \ngc{078},
 \ngc{7099} and IC$\,$4499. The collected data belong to the updated
2013 catalog of \citep{clement01} with few exceptions, namely \ngc{6121}
(Stetson et al. 2014, submitted), \ngc{6934} \citep{kaluzny01} and
\ngc{7006} \citep{wehlau99}.}
\end{deluxetable}

%%%%%%%%%%%%%%%%%%%%%%%%%%%%%%%%%%%%%%%%%%%%%%%%%%%%%%%%%%%%%%%%%%%%%%%%%%
%TAB 8
%%%%%%%%%%%%%%%%%%%%%%%%%%%%%%%%%%%%%%%%%%%%%%%%%%%%%%%%%%%%%%%%%%%%%%%%%%
\begin{deluxetable}{lccccccc}
%\centering
\tabletypesize{\tiny}
\tablewidth{0pt}
\tablenum{8}
\tablecaption{Chi-squared likelihood that RRab period distributions come from
  the same parent population (expressed in per cent).}
\label{tabprob}
\tablehead{
\colhead{Sample}    &
 \colhead{Inner halo a}&
 \colhead{Inner halo b}&
 \colhead{Outer halo a}&
 \colhead{Outer halo b}&
 \colhead{GCs} &
 \colhead{dSphs \& UFDs}&
 \colhead{LMC} \\
}
\startdata
Inner halo a&   100    &25(36)$^a$   &    0     &     0     &    5(2)$^a$    &       0     &   0 \\     
Inner halo b&   $-$    &    100     &    0     &     0     &     0(2)$^a$   &       0     &   0\\      
Outer halo a&   $-$    &    $-$     &   100    &     0     &       0      &       0     &   0\\      
Outer halo b&   $-$    &    $-$     &   $-$    &    100    &       0      &       0     &   0\\          
GCs         &   $-$    &    $-$     &   $-$    &    $-$    &      100     &       0     &   0\\   
dSphs \& UFDs&   $-$    &    $-$     &   $-$    &    $-$    &     $-$      &      100    &   0\\   
LMC         &   $-$    &    $-$     &   $-$    &    $-$    &     $-$      &      $-$    &  100 
\enddata 
\tablecomments{
$^a$ In parentheses we report the likelihood coming out of the
KS-test.
}
\end{deluxetable}

\clearpage
%%%%%%%%%%%%%%%%%%%%%%%%%%%%%%%%%%%%%%%%%%%%%%%%%%%%%%%%%%%%%%%%%%%%%%%%%%
%TAB 9
%%%%%%%%%%%%%%%%%%%%%%%%%%%%%%%%%%%%%%%%%%%%%%%%%%%%%%%%%%%%%%%%%%%%%%%%%%

\begin{deluxetable}{lllccllll}
\centering
\tabletypesize{\tiny}
\tablewidth{0pt}
\tablenum{9}
\tablecaption{Candidates from \cite{hodgewright78} that do not appear variable in our data}
\label{hodgewright}
\tablehead{
\colhead{ID} &
\colhead{$\left<B\right>$} &
\colhead{$\left<V\right>$} &
\colhead{$\left<R\right>$} &
\colhead{$\left<I\right>$} &
\colhead{RA} &
\colhead{Dec}
\\
\colhead{} &
\colhead{mag} &
\colhead{mag} &
\colhead{mag} &
\colhead{mag} &
\colhead{hh mm ss} &
\colhead{dd mm ss}
}
\startdata
HW4  & 23.72 & 22.95 & 22.52 & 22.02 & 10 08 32.79 & +12 18 37.6 \\
HW5  & 23.82 & 23.28 & 22.90 & 22.39 & 10 08 32.13 & +12 18 03.7 \\
HW6  & 21.55 & 20.64 & 20.09 & 19.55 & 10 08 37.97 & +12 17 45.5 \\
HW7  & 22.06 & 21.21 & 20.67 & 20.14 & 10 08 31.88 & +12 16 40.4 \\
HW9  & 21.53 & 20.29 & 19.58 & 18.95 & 10 08 28.36 & +12 16 04.5 \\
HW11 & 21.28 & 19.86 & 19.16 & 18.57 & 10 08 36.39 & +12 16 02.1 \\
HW12 & 22.46 & 21.04 & 20.43 & 19.91 & 10 08 40.51 & +12 15 04.5 \\
HW13 & 22.98 & 21.71 & 20.79 & 20.08 & 10 08 29.62 & +12 15 07.8 \\
HW15 & 21.81 & 20.88 & 20.32 & 19.79 & 10 08 27.98 & +12 16 53.5 \\
HW18 & 21.76 & 20.85 & 20.30 & 19.78 & 10 08 28.77 & +12 15 06.5 \\
HW20 & 20.81 & 19.21 & 18.27 & 17.54 & 10 08 15.86 & +12 20 32.6 \\
HW21 & 21.31 & 20.39 & 19.69 & 19.12 & 10 08 19.45 & +12 16 34.1 \\
HW22 & 20.90 & 19.75 & 19.05 & 18.41 & 10 08 26.48 & +12 18 23.9

\enddata
\end{deluxetable}

%%%%%%%%%%%%%%%%%%%%%%%%%%%%%%%%%%%%%%%%%%%%%%%%%%%%%%%%%%%%%%%%%%%%%%%%%%
%TAB 10
%%%%%%%%%%%%%%%%%%%%%%%%%%%%%%%%%%%%%%%%%%%%%%%%%%%%%%%%%%%%%%%%%%%%%%%%%%

\begin{deluxetable}{clcccccc}
\centering
\tabletypesize{\tiny}
\tablewidth{0pt}
\tablenum{10}
\tablecaption{Cross-identifications for stars from \cite{alw86}}
\label{alw}
\tablehead{
\colhead{ALW} &
\colhead{Our} &
\colhead{$B$} &
\colhead{$V$} &
\colhead{$R$} &
\colhead{$I$} &
\colhead{RA} &
\colhead{Dec}
\\
\colhead{ID} &
\colhead{ID\tablenotemark{a}} &&&&&
\colhead{hh mm ss} &
\colhead{dd mm ss}
\\
}
\startdata
 9 & \ $\,$91206  &  21.73 & 19.82 & 19.28 & 17.73 & 10 08 27.52 & +12 16 54.0 \\
10 & \ $\,$91627  &  21.21 & 19.45 & 18.42 & 17.62 & 10 08 27.69 & +12 17 31.9 \\
12 & \ $\,$97647  &  21.13 & 19.60 & 18.78 & 18.18 & 10 08 30.09 & +12 18 34.3 \\
14 & \ $\,$99852  &  20.70 & 18.83 & 17.83 & 17.03 & 10 08 31.04 & +12 17 07.8 \\
15 & \ $\,$67567$\,=\,$V79  &  21.34 & 19.28 & 18.15 & 17.16 & 10 08 34.72 & +12 18 37.7 \\
18 & 109932 &  21.18 & 19.28 & 18.33 & 17.59 & 10 08 35.30 & +12 17 24.6 \\
19 & \ $\,$62888  &  21.12 & 19.31 & 18.41 & 17.72 & 10 08 16.61 & +12 20 07.2 \\
20 & \ $\,$66463  &  21.64 & 19.49 & 18.40 & 17.50 & 10 08 18.09 & +12 20 09.4 \\
\enddata
\tablenotetext{a}{Sequential identification number in our catalog for \leoi, available on-line.}
\end{deluxetable}

\clearpage

\clearpage

\clearpage

\clearpage

\clearpage

\clearpage

\clearpage

\clearpage

\clearpage

\clearpage

\clearpage

\clearpage

\clearpage

\end{document}